\documentclass[11pt]{article}
\pdfoutput=1
\usepackage{soul}
\usepackage{array}
\usepackage{jheppub}
\usepackage{amsfonts}
\usepackage{amsmath}
\allowdisplaybreaks[4]         
\usepackage{amssymb}   
\usepackage{euscript}       
\usepackage{xcolor}          
\usepackage{tensor}     
\usepackage{caption}
\usepackage{subcaption}   
\usepackage{graphicx}
\usepackage[T1]{fontenc} 
\usepackage{float}

\newcommand{\req}[1]{(\ref{#1})} 

\newcommand{\bea}{\begin{eqnarray}}
	\newcommand{\eea}{\end{eqnarray}}
\newcommand{\ba}{\begin{eqnarray}}
	\usepackage{braket}
	\newcommand{\ea}{\end{eqnarray}}

\newcommand{\beq}{\begin{equation}}
	\newcommand{\eeq}{\end{equation} }
\newcommand{\beqa}{\begin{eqnarray}}
	
	\newcommand{\eeqa}{\end{eqnarray}}
\newcommand{\beqar}{\begin{eqnarray*}}
	\newcommand{\eeqar}{\end{eqnarray*}}

\newcommand{\be}{\begin{equation}}
	\newcommand{\ee}{\end{equation}}

\renewcommand{\req}[1]{(\ref{#1})}

\newcommand{\A}{\mathcal{A}}
\newcommand{\B}{\mathcal{B}}

\newcommand{\E}{\mathcal{E}}

\newcommand{\thickhline}{%
    \noalign {\ifnum 0=`}\fi \hrule height 1pt
    \futurelet \reserved@a \@xhline
}
\newcolumntype{"}{@{\hskip\tabcolsep\vrule width 1pt\hskip\tabcolsep}}
\makeatother

\allowdisplaybreaks

\title{Love numbers beyond GR from the modified Teukolsky equation}

\author[a]{Pablo A. Cano}

\affiliation[a]{Departament de F\'isica Qu\`antica i Astrof\'isica, Institut de Ci\`encies del Cosmos Universitat de Barcelona, Mart\'i i Franqu\`es 1, E-08028 Barcelona, Spain}

\emailAdd{pablo.cano@icc.ub.edu}

\date{\today}
\abstract{We obtain the full set of tidal Love numbers of non-rotating black holes in an effective field theory extension of general relativity. We achieve our results using a recently introduced modified Teukolsky equation that describes the perturbations of black holes in this theory. 
We show how to identify the Love numbers and their beta functions in a systematic and gauge invariant way, applying analytic continuation on the angular number $\ell$ when necessary.
We observe that there are three types of Love numbers: electric, magnetic, and a ``mixing'' type, associated to parity-breaking theories, that we identify here for the first time. 
The modified Teukolsky equation proves to be very useful as it allows us to obtain all the different Love numbers in a unified framework. 
We compare our results with previous literature that utilized the Regge-Wheeler-Zerilli equations to compute Love numbers, finding perfect agreement. 
The method introduced here paves the way towards the computation of Love numbers of rotating black holes beyond general relativity. 
}

\begin{document} 
	\maketitle
	\flushbottom

	\newpage
	\allowdisplaybreaks
	
\section{Introduction}
Tidal Love numbers (TLNs) characterize the deformation of a body due to an external tidal field, and are nowadays a subject of interest in the context of general relativity (GR) \cite{Binnington:2009bb}.
The study of tidal properties of black holes has spiked in recent years, to a big extent due to its relevance for gravitational wave observations of compact binaries \cite{LIGOScientific:2016aoc}. Tidal deformability of each of the objects in a compact binary affects the gravitational wave emission during the inspiral \cite{Damour:2009vw,Vines:2011ud,Bini:2012gu}, so TLNs are relevant quantities to correctly model the evolution of a binary and to test the nature of the compact objects \cite{Cardoso:2017cfl}.

A famous result is that black holes in four-dimensional vacuum GR have vanishing Love numbers, and therefore are not deformed by tidal fields \cite{Binnington:2009bb,Kol:2011vg,Hui:2020xxx,LeTiec:2020bos,Charalambous:2021mea,Charalambous:2021kcz,Ivanov:2022qqt}. 
Thus, if we experimentally observed a non-zero tidal deformation, this would imply that either (i) GR is wrong/incomplete, or (ii) the objects detected are not black holes. 
Therefore, it is interesting to investigate extensions of Einstein's theory, as some of them can lead to non-zero Love numbers.  In this direction, there has been important progress in the analysis of TLNs of non-rotating black holes in a number of theories beyond GR \cite{Cardoso:2017cfl,Cardoso:2018ptl,Chakravarti:2018vlt,Cai:2019npx,DeLuca:2022tkm,Katagiri:2023umb,Katagiri:2024fpn,Chakraborty:2024gcr,Barbosa:2025uau}, including also the development of a parametrized formalism in \cite{Katagiri:2023umb}.\footnote{See also \textit{e.g.} \cite{Yazadjiev:2018xxk,Diedrichs:2025vhv} for TLNs of neutron stars in theories beyond GR.} These works rely on the Regge-Wheeler-Zerilli approach \cite{regge1957stability,zerilli1970effective} to analyze black hole perturbations. This involves a decomposition of the metric perturbation in spherical harmonics followed by a reduction of the linearized equations into one master equation for axial perturbations (Regge-Wheeler) and another one for polar perturbations (Zerilli).  Different techniques are then employed in order to extract the corresponding TLNs. Despite this progress, we note that available results are not fully exhaustive: TLNs are usually obtained only for a few multipoles $\ell$, and some theories, like those that break parity, remain to be investigated. 
In fact, it was noted in \cite{Cardoso:2018ptl} that parity-violating theories would introduce new types of Love numbers, but these have not been studied yet. 

The case of rotating black holes is much more challenging, and to the best of our knowledge, TLNs of rotating black holes in extensions of GR have not ever been computed. In fact, the analysis of TLNs of Kerr black holes in GR is an active topic of research \cite{LeTiec:2020spy,LeTiec:2020bos,Chia:2020yla,Charalambous:2021mea,Rodriguez:2023xjd,Perry:2023wmm,Perry:2024vwz,Bhatt:2024yyz}. 
These analyses are made possible by the Teukolsky equation \cite{Teukolsky:1972my,Teukolsky:1973ha} --- a master equation for curvature perturbations on the Kerr background in GR. 
Generalized Teukolsky equations describing the perturbations of rotating black holes in extensions of GR have been recently developed by \cite{Li:2022pcy,Hussain:2022ins,Cano:2023tmv,Cano:2023jbk,Wagle:2023fwl,Cano:2024ezp,Guo:2024bqe}. 
These modified Teukolsky equations should allow us to investigate the TLNs of rotating black holes. However, we have to address the challenge of how to identify the Love numbers from those equations.  This is a non-trivial problem since the usual definition of Love numbers from black hole perturbation theory possesses certain ambiguities \cite{Gralla:2017djj,LeTiec:2020bos,Charalambous:2021mea,Ivanov:2022hlo,Bhatt:2023zsy}.  Let us briefly recall the main obstacles. 

\subsection{The problem of defining Love numbers}
In order to give an intuitive idea of TLNs, we start by recalling their definition in Newtonian gravity. We consider a spherically symmetric body of mass $M$. We then apply an external $\ell$-polar tidal field, which in the near zone of the body is given by $U^{\rm tidal}=\mathcal{E}_{i_{1}\ldots i_{\ell}} x^{i_{1}}\ldots x^{i_{\ell}}\equiv \mathcal{E}_{L}n^{L} r^{\ell}$, where $\mathcal{E}_{i_{1}\ldots i_{\ell}}$ is a symmetric and traceless tensor. We have introduced the notation $\mathcal{E}_{L}n^{L}=\mathcal{E}_{i_{1}\ldots i_{\ell}}n^{i_1}\ldots n^{i_{\ell}}$, where $n^{i}=x^{i}/r$ is the unit vector. The body will react to the external field and will generate induced multipole moments $M_{L}$. The Love numbers $k_{\ell}$ are nothing but the proportionally constant between $M_{L}$ and the external field $\mathcal{E}_{L}$, so that the total Newtonian potential reads
\begin{equation}\label{eq:Newt}
U=-\frac{M}{r}+\sum_{\ell=2}\mathcal{E}_{L} n^{L} r^{\ell}\left[1-2k_{\ell}\left(\frac{R_{o}}{2r}\right)^{2\ell+1}\right]\, ,
\end{equation}
where $R_o$ is the radius of the object, which we have introduced to make $k_{\ell}$ dimensionless. Therefore, the TLN can be simply identified by looking at the coefficient of $r^{-\ell-1}$ in the Newtonian potential. While this definition is fine in Newtonian gravity, its relativistic generalization possesses several ambiguities. The GR version of \req{eq:Newt} includes relativistic corrections to both the source term $r^{\ell}$ and to the response term $r^{-\ell-1}$. Schematically, we get an expression of the form \cite{Kol:2011vg} 
\begin{equation}\label{eq:NewtGR}
U^{\rm GR}=-\frac{M}{r}+\sum_{\ell=2}\mathcal{E}_{L} n^{L} r^{\ell}\left[\left(1+\sum_{n=1}a_{n}\frac{R_{o}^{n}}{r^n}\right)-2k_{\ell}\left(\frac{R_{o}}{2r}\right)^{2\ell+1}\left(1+\sum_{n=1}b_{n}\frac{R_{o}^{n}}{r^n}\right)\right]\, ,
\end{equation}
where the coefficients $a_n$, $b_n$ represent the relativistic corrections and $U^{\rm GR}=-(g_{00}+1)/2$ in an appropriate gauge. Thus, the coefficient of $r^{-\ell-1}$ contains in general a contribution from relativistic corrections (the coefficient $a_{2\ell+1}$) besides the Love number $k_{\ell}$. Therefore, it is not possible to  distinguish the response from the tidal field and we cannot unambiguously identify the TLNs. 
On the other hand, one may just define the TLN as the coefficient of $r^{-\ell-1}$, but this is a coordinate-dependent definition \cite{Gralla:2017djj}.

A rigorous way of defining tidal deformability is through the construction of an effective field theory (EFT) for the worldline action of a point particle, representing a body (in our case a black hole) seen from far away. In this set-up, it is possible to  supplement the worldline action with operators that account for tidally-induced multipole moments of the body  \cite{Goldberger:2004jt,Porto:2005ac,Levi:2018nxp}. These operators are weighted by tidal coefficients $\lambda_{\rm \ell}$ that are well-defined and unambiguous. In order to obtain the value of such coefficients, one needs to match the predictions of the worldline EFT with the results of black hole perturbation theory in the near zone. As nicely explained in \cite{Ivanov:2022hlo}, this matching should be performed in the same gauge and accounting for all the relativistic corrections.

However, applying this technique to the case of modified gravity is beyond our current capabilities. For instance, in the context of the modified Teukolsky equation, the meaning of the Teukolsky variable is obscure (many changes of variable are implemented to simplify the form of the equation) and it is not currently known how to reconstruct the full metric perturbation in terms of the Teukolsky variable. Therefore, we do not have all the information required to perform an exact matching with the worldline EFT. 

To circumvent these issues, we would need to have a robust notion of TLNs in black hole perturbation theory, \textit{i.e.}, a notion that is invariant under coordinate changes and redefinitions of variable. 
Love numbers obtained through a gauge-invariant prescription should be, up to a normalization factor, the tidal coefficients of the worldline EFT.

A popular technique to define TLNs in a gauge-invariant way is to perform an analytic continuation in the harmonic number $\ell$, allowing it to take real, or even complex values. The idea is that, if $\ell$ is an arbitrary number, then the $r^{\ell}$ and $r^{-\ell-1}$ series in \req{eq:NewtGR} do not overlap, and one can unambiguously read-off the TLN $k_{\ell}$. Furthermore, this identification is robust and invariant under coordinate transformations and changes of variable, as we explain in more detail in section \ref{sec:id}. 
This technique applied to GR predicts the vanishing of the TLNs of Schwarzschild \cite{Binnington:2009bb,Kol:2011vg,Hui:2020xxx} and Kerr black holes \cite{LeTiec:2020bos,Chia:2020yla,Charalambous:2021mea}, and it is consistent with the conclusions obtained from scattering amplitudes in the worldline EFT \cite{Ivanov:2022hlo,Ivanov:2022qqt,Saketh:2023bul}.  Therefore, the TLNs defined via analytic continuation in $\ell$ should be directly related to the tidal coefficients of the worldline EFT --- see \textit{e.g.} \cite{Hui:2020xxx}. 

In this paper, we apply these ideas to the modified Teukolsky equation in order to obtain TLNs of non-rotating black holes in higher-derivative gravity, in the form of an EFT extension of GR.\footnote{This is an EFT for the gravitational field and should not be confused with the worldline EFT, which is an effective theory for a gravitating particle.}  
The application of analytic continuation is not straightforward, as it requires that we obtain the solution of the modified Teukolsky equation analytically in $\ell$, which is not always possible. We introduce a method that enables us to perform such computation, by expanding the solution near integer values of $\ell$. 
Our analysis reveals that the use of analytic continuation is crucial to obtain the correct result; we reproduce in this way previous results obtained through the Regge-Wheeler-Zerilli approach \cite{Cardoso:2018ptl,Cai:2019npx,DeLuca:2022tkm,Katagiri:2023umb,Katagiri:2024fpn,Barbosa:2025uau}. The Teukolsky approach turns out to be very powerful, as it allows us to characterize the Love numbers in terms of a single theory-dependent coefficient that enters in the modified Teukolsky equation.  Furthermore, it also allows us to characterize the different types of TLNs --- associated to the parity of the tidal field --- in a unified way. These include the electric and magnetic type Love numbers, as well as a new type of Love number introduced by parity-violating theories that we identify here for the first time. 

The rest of the paper is organized as follows
\begin{itemize}
\item In section~\ref{sec:EFT} we review the EFT extension of GR, its spherically symmetric black hole solutions, and the modified Teukolsky equation derived in \cite{Cano:2023tmv,Cano:2023jbk,Cano:2024ezp}. We reduce this equation in the case of static perturbations and show that it takes a simple form involving only two theory-dependent coefficients.  
\item In section~\ref{sec:id} we analyze the solutions of the static modified Teukolsky equation and we identify the Love numbers. We find that, in general, the asymptotic expansion of the solution includes logarithmic terms, which can be interpreted as a running of the Love numbers in the context of the worldline EFT \cite{Kol:2011vg,Charalambous:2021mea,Ivanov:2022hlo,Barbosa:2025uau}.  The coefficient of the logarithm is identified with a beta function and we show that its value is unambiguous. In the cases when the beta functions vanish, we use analytic continuation in order to extract the (constant) Love numbers. We present results for generic $\ell$ in terms of a single coefficient entering in the modified Teukolsky equation. 
\item In section~\ref{sec:loveEFT} we apply the results of the preceding section in order to obtain the explicit values of TLNs and their beta functions in the EFT of GR. We show that there are three types of Love numbers: electric (polar response to a polar perturbation), magnetic (axial response to an axial perturbation) and a new ``mixing'' type (polar response to an axial perturbation, and viceversa). We show how to identify each type of TLN from the results of the modified Teukolsky equation, and in particular, we find that parity-violating corrections induce mixing TLNs. 
We give complete results for all the beta functions and all the non-running Love numbers and we compare them with the partial results in previous literature, finding perfect agreement. 
\item We conclude in section~\ref{sec:conclusions}, where we further discuss our results as well as future prospects. 
\end{itemize} 

\section{EFT of GR: black holes and perturbation theory}\label{sec:EFT}
To eight-derivative order, a general EFT extension of the Einstein-Hilbert action can be expressed as a combination of two cubic and three quartic curvature invariants as \cite{Endlich:2017tqa, Cano:2019ore}
\begin{equation}\label{eq:EFT}
\begin{aligned}
S_{\rm EFT}=\frac{1}{16\pi G}\int d^4x\sqrt{|g|}\bigg[R&+\lambda_{\rm ev}\tensor{R}{_{\mu\nu }^{\rho\sigma}}\tensor{R}{_{\rho\sigma }^{\delta\gamma }}\tensor{R}{_{\delta\gamma }^{\mu\nu }}+\lambda_{\rm odd}\tensor{R}{_{\mu\nu }^{\rho\sigma}}\tensor{R}{_{\rho\sigma }^{\delta\gamma }} \tensor{\tilde R}{_{\delta\gamma }^{\mu\nu }}\\
&+\xi_{1}(\mathcal{R}^2)^2+\xi_{2}(\tilde{\mathcal{R}}^2)^2+\xi_{3}\mathcal{R}^2\tilde{\mathcal{R}}^2+\ldots \bigg]\, ,
\end{aligned}
\end{equation}
where
\begin{align}
\mathcal{R}^2&=R_{\mu\nu\rho\sigma} R^{\mu\nu\rho\sigma}\, ,\quad 
\tilde{\mathcal{R}}^2=R_{\mu\nu\rho\sigma} \tilde{R}^{\mu\nu\rho\sigma}\, , 
\end{align}
and where
\begin{equation}
 \tilde{R}_{\mu\nu\rho\sigma}=\frac{1}{2}\epsilon_{\mu\nu\alpha\beta}\tensor{R}{^{\alpha\beta}_{\rho\sigma}}
 \end{equation}
 is the dual Riemann tensor.  The couplings are dimensionful and their scale is related to a length scale of new physics $L_{\rm new}$ so that  $\lambda_{\rm ev, odd}\sim L_{\rm new}^{4}$, $\xi_{i}\sim L_{\rm new}^{6}$. The terms proportional to $\lambda_{\rm ev}$ and $\xi_{3}$ break parity, and we will see that this has interesting consequences for the tidal deformability of black holes.
 
 The equations of motion read 
 \begin{align}\label{EFE}
\E_{\mu\nu}\equiv  G_{\mu\nu}+\tensor{P}{_{(\mu}^{\rho \sigma \gamma}} R_{\nu) \rho \sigma \gamma}+2 \nabla^\sigma \nabla^\rho P_{(\mu|\sigma| \nu) \rho}-\frac{1}{2} g_{\mu \nu} \mathcal{L}^{\rm HD}=0\,,
\end{align}
where $\mathcal{L}^{\rm HD}=\lambda_{\rm ev}\tensor{R}{_{\mu\nu }^{\rho\sigma}}\tensor{R}{_{\rho\sigma }^{\delta\gamma }}\tensor{R}{_{\delta\gamma }^{\mu\nu }}+\ldots$ is the higher-derivative part of the Lagrangian and $ P_{\mu \nu \rho \sigma}$ is the derivative of $\mathcal{L}^{\rm HD}$ with respect to the Riemann tensor, given by
\begin{align}\notag
	 P_{\mu \nu \rho \sigma}&=3 \lambda_{\mathrm{ev}} R_{\mu \nu}{}^{\alpha \beta} R_{\alpha \beta \rho \sigma}+\lambda_{\text {odd }}\left(R_{\mu \nu}{ }^{\alpha \beta} \tilde{R}_{\alpha \beta \rho \sigma}+R_{\mu \nu}{ }^{\alpha \beta} \tilde{R}_{\rho \sigma \alpha \beta}+R_{\rho\sigma}{ }^{\alpha \beta} \tilde{R}_{\mu\nu \alpha \beta}\right)\\
	 &+4 \xi_1 \mathcal{R}^2 R_{\mu \nu \rho \sigma}+2 \xi_2 \tilde{\mathcal{R}}^2\left(\tilde{R}_{\mu \nu \rho \sigma}+\tilde{R}_{\rho \sigma \mu \nu}\right)+\xi_3\left[2 \tilde{\mathcal{R}}^2 R_{\mu \nu \rho \sigma}+\mathcal{R}^2\left(\tilde{R}_{\mu \nu \rho \sigma}+\tilde{R}_{\rho \sigma \mu \nu}\right)\right]\,.
\end{align}
In what follows, we treat the higher-derivative corrections as a perturbative expansion around GR, to first order in the coupling constants. We work in units of $G=1$ from now on. 

\subsection{Black hole solutions}
The static black hole solutions of \req{eq:EFT} can be written as a modification of the Schwarzschild metric as follows
\begin{equation}\label{correctedmetric}
ds^2=-\left(1+h_1(r)\right) f(r) dt^2+\left(1+h_2(r)\right)\left[\frac{dr^2}{f(r)}+r^2\left(d\theta^2+\sin^2\theta d\phi^2\right)\right]\, ,
\end{equation}
where 
\begin{equation}
f(r)=1-\frac{2M}{r}
\end{equation}
and the functions $h_1$, $h_2$ capture the corrections. At linear order in the coupling constants, they read
\begin{equation}
\begin{aligned}
h_{1}&=\frac{\lambda_{\rm ev}}{M^4} \left(\frac{64}{231x}+\frac{64}{231 x^2}+\frac{32}{77 x^3}+\frac{160}{231 x^4}+\frac{40}{33 x^5}+\frac{24}{11x^6}\right)\\
&+\frac{\xi_{1}}{M^6} \left(\frac{16384}{12155 x}+\frac{16384}{12155 x^2}+\frac{24576}{12155 x^3}+\frac{8192}{2431 x^4}+\frac{14336}{2431
   x^5}+\frac{129024}{12155 x^6}+\frac{21504}{1105 x^7}\right.\\
   &\left.+\frac{3072}{85 x^8}-\frac{16256}{17 r^9}\right)\, ,\\
h_{2}&=\frac{\lambda_{\rm ev}}{M^4}\left(-\frac{64}{231}+\frac{64}{231
   x}+\frac{32}{231 x^2}+\frac{32}{231 x^3}+\frac{40}{231 x^4}+\frac{8}{33 x^5}-\frac{392}{11 x^6}\right)\\
   &+\frac{\xi_{1}}{M^6} \left(-\frac{16384}{12155}+\frac{16384}{12155
   x}+\frac{8192}{12155 x^2}+\frac{8192}{12155 x^3}+\frac{2048}{2431 x^4}+\frac{14336}{12155
   x^5}\right.\\
   &\left.+\frac{21504}{12155 x^6}+\frac{3072}{1105 x^7}+\frac{384}{85 x^8}-\frac{8576}{17 x^9}\right)\, ,
\end{aligned}
\end{equation}
where $x=r/M$. Although this way of expressing the corrections may not be the most compact one, the ansatz \req{correctedmetric} has the advantage that it keeps the position of the horizon fixed at $r=2M$.\footnote{Nevertheless, the properties of the horizon do change,  like for instance its area, $A=16\pi M^2(1+h_{2}(2M))$. }  This is convenient for the analysis of perturbations, as one does not need to worry about the position of the horizon --- which is a singular point of the perturbation equations --- receiving corrections.  On the other hand, this form of the metric is directly generalizable to the case of rotating black holes by using the ansatz of \cite{Cano:2019ore}. In fact, we have obtained \req{correctedmetric} by setting the angular momentum to zero in the solutions obtained in \cite{Cano:2019ore} and \cite{Cano:2020cao}. 

\subsection{The modified Teukolsky equation}\label{sec:modifiedTeuk}
The perturbations of the static black holes \req{correctedmetric} can be studied through a direct decomposition of the metric perturbation in spherical harmonics, which gives rise to modified Regge-Wheeler \cite{regge1957stability} and Zerilli \cite{zerilli1970effective} master equations. This analysis has been carried out in a detailed way in \cite{Cardoso:2018ptl,deRham:2020ejn,Cano:2021myl}, among others. 

However, here we will make use of the Teukolsky approach to black hole perturbation theory. There are two reasons why this approach is advantageous. First, as our analysis below shows, the modified Teukolsky equation allows us to describe all types of perturbations (axial/polar) and all of the higher-derivatives corrections in \req{eq:EFT} in a unified way. Second and most important, it admits a direct generalization to the case of rotating black holes. 

Modified Teukolsky equations for extensions of GR have been introduced by Refs.~\cite{Li:2022pcy,Hussain:2022ins,Cano:2023tmv}. Here we will use the results of  \cite{Cano:2023tmv,Cano:2023jbk,Cano:2024ezp} which have found the explicit form of modified radial Teukolsky equations that govern the perturbations of rotating black holes in the EFT \req{eq:EFT}.  It is straightforward to reduce these equations to the case of static black holes by setting the rotation parameter to zero, so we refer to those papers for details about the derivation of these equations. 

There are two Teukolsky equations for gravitational perturbations: one of spin $s=+2$ and another one of spin $s=-2$, associated to the Teukolsky variables $\Psi_{0}$ and $\Psi_{4}$, respectively.  The two equations are equivalent --- they are related by a transformation --- so we focus on the equation of spin $s=-2$. The fluctuations of $\Psi_{4}$ over the background black hole geometry are decomposed as
\begin{equation}
\delta\Psi_{4}=e^{-i\omega t+i m\phi} r^{-4} S_{-2}^{\ell m}(\theta) \psi(r)\, ,
\end{equation}
where $S^{\ell m}_{-2}(\theta)$ are the spin-weighted spherical harmonics of spin $s=-2$ and $\psi(r)$ is the radial variable, whose indices $\ell$, $m$, $s$ are omitted to streamline the notation. The master equation for the radial variable $\psi(r)$ reads
\begin{equation}\label{Teuk}
\Delta^{2}\frac{d}{dr}\left(\Delta^{-1}\frac{d\psi}{dr}\right)+\left(V_{\ell}+\alpha \delta V_{\ell}\right)\psi=0\, ,
\end{equation}
where 
\begin{equation}
\Delta=r(r-2M)\, ,
\end{equation}
 $V_{\ell}$ is the Teukolsky potential
\begin{equation}
V_{\ell}=\frac{\omega^2r^4-4i r^2 (r-3M)\omega}{\Delta}+2-\ell-\ell^2 \, ,
\end{equation}
and $\delta V_{\ell}$ is a correction to the potential due to the higher-derivative terms. The parameter $\alpha$ is a bookkeeping parameter that we use to control the higher-derivative expansion, but we will set it to 1 at the end of the computations, since  $\delta V_{\ell}$ is already a combination of the different couplings $\lambda_{\rm ev, odd}$, $\xi_{i}$.  The form of this potential can be changed by perturbative redefinitions of the radial variable
\begin{equation}\label{Psitrans}
\psi\to \psi+\alpha\left( f_{1}(r)\psi+f_{2}(r) \Delta \frac{d\psi}{dr}\right)\, ,
\end{equation}
which are equivalent to a redefinition of the background Newman-Penrose frame in which we compute $\Psi_{4}$. The functions $f_1(r)$ and $f_{2}(r)$ must be smooth and fall-off at infinity fast enough, but are otherwise arbitrary.  Physical quantities like quasinormal modes and Love numbers are not affected by this type of transformations, as long as our definitions are gauge-invariant --- we discuss this in section~\ref{sec:id}.  It was shown in \cite{Cano:2024ezp} that, by applying appropriate redefinitions of the form \req{Psitrans}, it is always possible to put the correction to the potential into the form\footnote{Observe that here $\delta V_{\ell}$ is defined with an extra factor of $1/\Delta$ with respect to the definition of \cite{Cano:2023tmv,Cano:2023jbk,Cano:2024ezp}.}
\begin{equation}\label{eq:deltaV}
\delta V_{\ell}=\frac{M^2}{\Delta}\left[A_{-2,\ell}(\omega)\left(\frac{M}{r}\right)^2+A_{0,\ell}(\omega)+A_{1,\ell}(\omega)\left(\frac{r}{M}\right)+A_{2,\ell}(\omega) \left(\frac{r}{M}\right)^2\right]\, ,
\end{equation}
where the coefficients $A_{k,\ell}(\omega)$ are functions of $\omega$ and $\ell$.  Importantly, these coefficients also depend on an additional parameter $q_{-2}$ that is related to the polarization of the perturbation. This does not appear in the GR Teukolsky equation due to isospectrality, but it appears in extensions of GR. We omit the dependence on $q_{-2}$ to shorten the notation, but this parameter will be important for us in section \ref{sec:polarization}.  Obtaining \req{eq:deltaV} involves a very long and technical computation following the procedure developed in \cite{Cano:2023tmv,Cano:2023jbk,Cano:2024ezp}. Here we have used the techniques detailed in those references to obtain $\delta V_{\ell}$ and the coefficients $A_{k,\ell}(\omega)$ for arbitrary $\ell$ in the case of non-rotating black holes. The explicit expressions for these coefficients are long and they can be found in the repository \cite{gitbeyondkerr}.

Since our interest is on static Love numbers, from now on we restrict ourselves to static perturbations $\omega=0$. In this case, we observe that the form of \req{eq:deltaV} simplifies substantially. Using the explicit values of the $A_{k,\ell}(\omega)$ coefficients, we observe the relationships
\begin{equation}
A_{2,\ell}(0)=0\, ,\quad \frac{1}{4}A_{-2,\ell}(0)+A_{0,\ell}(0)+2A_{1,\ell}(0)=0\, .
\end{equation}
This allows us to express $\delta V_{\ell}$ using only two coefficients 
\begin{equation}\label{dVom0}
\delta V_{\ell}\big|_{\omega=0}=B_\ell\frac{M}{ r}+C_{\ell}\frac{M^2\left(r+2M\right)}{r^3}\, ,
\end{equation}
where $B_{\ell}= A_{1,\ell}(0)$ and $C_{\ell}=-A_{-2,\ell}(0)/4$.  We show the coefficients $B_{\ell}$, $C_{\ell}$ for each of the higher-derivative corrections Appendix~\ref{app:coef}. Crucially, we see that for static perturbations the potential is finite at the horizon, so the nature of the regular singular point at the horizon is not modified. 

Putting all the pieces together, the differential equation reads, explicitly
\begin{equation}\label{Teuk2}
\Delta\frac{d^2\psi}{dr^2}-2(r-M)\frac{d\psi}{dr}+\left[2-\ell-\ell^2+\alpha\left(B_\ell\frac{M}{ r}+C_{\ell}\frac{M^2\left(r+2M\right)}{r^3}\right)\right]\psi=0\, .
\end{equation}
To unveil the nature of the singular points we perform the change of variables $r=2M/z$, and we get
\begin{equation}
z^2(1-z)\frac{d^2\psi}{dz^2}+z(4-3z)\frac{d\psi}{dz}+\left[2-\ell-\ell^2+\alpha\left(B_{\ell}\frac{z}{2}+C_{\ell}\frac{z^2\left(1+z\right)}{4}\right)\right]\psi=0\, .
\end{equation}
It follows that both the horizon $z=1$ and infinity $z=0$ are regular singular points. 

Finally, let us point out that, just like \req{Teuk2} is the modified Teukolsky equation for the $\Psi_{4}$ variable, one can find the corresponding equation for the Newman-Penrose conjugate variable $\Psi_{4}^{*}$.\footnote{The Newman-Penrose conjugation of a quantity $X$ is defined as the quantity obtained by the exchange $m_{\mu}\leftrightarrow \bar{m}_{\mu}$ of the two complex frame vectors and is denoted by $X^{*}$. It is in general not equivalent to complex conjugation, which is denoted by a bar $\bar{X}$, since we allow the metric perturbation to be complex.} We find, via an explicit computation, that this one takes the same form as \req{Teuk2} but with different coefficients $B^{*}_{\ell}$, $C^{*}_{\ell}$, which turn out to be related to $B_{\ell}$, $C_{\ell}$ by 
\begin{equation}\label{coefdual}
B^{*}_{\ell}(q_{-2})=\bar{B}_{\ell}(1/q_{-2})\, ,\quad C^{*}_{\ell}(q_{-2})=\bar{C}_{\ell}(1/q_{-2})\, ,
\end{equation}
where the complex conjugation only acts on explicit appearances of $i$, not on $q_{-2}$. We will come back to this in section \ref{sec:polarization}. For now, we focus on the equation \req{Teuk2}. 

\section{Identifying the Love numbers}\label{sec:id}
The equation \req{Teuk2} has a regular singular point at the horizon and another one at infinity. We observe that, since the corrections to the potential decay faster at infinity than the GR contribution, the solutions have the same asymptotic structure as in GR.  For $r\to\infty$, the solution consists of a series that starts with a power of $r^{\ell+2}$ and another one that starts at $r^{1-\ell}$: 
\begin{equation}\label{eq:asympt}
\psi(r)=\mathcal{A}\left(\frac{r}{M}\right)^{\ell+2}\left[1+a_1 \left(\frac{M}{r}\right)+\ldots\right]+\mathcal{B} \left(\frac{M}{r}\right)^{\ell-1}\left[1+b_1 \left(\frac{M}{r}\right)+\ldots\right]\, ,
\end{equation}
where $\mathcal{A}$ and $\mathcal{B}$ are free integration constants. 
At the horizon, there are again two independent solutions, but only one of them is regular and represents the physical solution.  Once we select the solution that is regular at the horizon, the ratio $\B/\A$ in the asymptotic expansion \req{eq:asympt} is fixed. One can interpret the leading term $\mathcal{A} r^{\ell+2}$ as an external tidal field, and $\mathcal{B} r^{1-\ell}$ as a response. Thus, we define the tidal Love numbers as
\begin{equation}\label{LoveDef}
k_{\ell}=\frac{\B}{2\A}\quad \text{for the regular solution}\, ,
\end{equation}
so that they measure the response due to an external tidal field. 

However, this definition suffers from several ambiguities. We note that the $\A$-series in \req{eq:asympt}, can in principle extend to arbitrary negative powers of $r$, so it may contain a term $r^{1-\ell}$ as well. Therefore, the coefficient of $r^{1-\ell}$ may contain not only the response coefficient $\B$, but also a contribution from the external field. Thus, from a given solution $\psi(r)$ one cannot unambiguously separate the tidal field from the response. 

On the other hand, one may simply define the Love number from the coefficient of $r^{1-\ell}$, without worrying if it comes from the response or from the external field. However, this is a gauge-dependent quantity, since this coefficient can be changed by coordinate transformations such as 
\begin{equation}\label{radialchange}
r\to r+c_1 M/r+c_2 (M/r)^2+\ldots\, ,
\end{equation}
and by redefinitions of $\psi$ like \req{Psitrans}. 
These problems are well known and have been pointed out repeatedly in the literature, \textit{e.g.} \cite{Gralla:2017djj,LeTiec:2020bos,Charalambous:2021mea,Ivanov:2022hlo}.

A simple trick to extract an unambiguous and gauge-invariant Love number is to perform an analytic continuation in the angular number $\ell$ by allowing it to take arbitrary real values and not only integer ones. We will use the notation $\hat{\ell}$ to refer to the analytically extended angular number, while we will use the unhatted $\ell$ for integer values, 
\begin{equation}
\hat{\ell}\in \mathbb{R}\, , \qquad \ell \in \mathbb{N}\, .
\end{equation}
When we write ``$\ell$'' we also have an in mind that $\ell$ takes a particular value. 

Observe that, if we were able to obtain \req{eq:asympt} analytically for an arbitrary real $\hat{\ell}$, then we could extract $\B$ and the Love number unambiguously, since the $r^{1-\hat{\ell}}$ and $r^{2+\hat{\ell}}$ series do not mix. Furthermore, the Love number identified in this way is a gauge-invariant quantity. It is easy to convince oneself that the ratio $\B/\A$ is not modified by changes of coordinates like \req{radialchange}, or redefinitions of variable  like \req{Psitrans}, as long as these transformations are smooth at infinity.\footnote{Contrarily as stated in \cite{LeTiec:2020bos}, one can allow for transformations of coordinates that depend on $\ell$. However, they must be smooth for arbitrary values of $\hat{\ell}$. For instance, a coordinate transformation like $r\to r+c_1 (M/r)^{\hat{\ell}}+c_2 (M/r)^{2\hat{\ell}}+\ldots$, which could affect the value of $k_{\hat{\ell}}$, is not allowed because it is non-smooth at infinity for real $\hat{\ell}$. However, a transformation like $r\to r+c_1(\hat{\ell}) (M/r)+c_2(\hat{\ell}) (M/r)^{2}+\ldots$ is smooth and hence acceptable, and it leaves invariant the value of $k_{\hat{\ell}}$.} Thus, the identification of Love numbers via analytic continuation is robust.

In some cases, the coefficient of $r^{1-\ell}$ may not be constant, as it may contain a logarithm, 
\begin{equation}\label{betadef}
\B=2\A \left[\beta_{\ell} \log\left(\frac{r}{M}\right)+k^{0}_{\ell}\right]\, .
\end{equation}
In these cases, the coefficient of the logarithm, $\beta_{\ell}$, is interpreted as the beta function of the Love number from the perspective of the worldline EFT \cite{Kol:2011vg,Charalambous:2021mea,Ivanov:2022hlo,Barbosa:2025uau}. If $r^{1-\ell}$ is the highest power of $r$ that comes accompanied by a logarithm, then $\beta_{\ell}$ is an invariant quantity and we do not need to resort to analytic continuation. To see this, we note that the only way to change the coefficient of this logarithm would be to perform coordinate changes such as $r\to r+c_{1}r^{-n}\log(r/M)+\ldots$, which are not allowed since they are not smooth at infinity (a similar comment applies to redefinitions of $\psi$). On the other hand, the constant part $k^{0}_{\ell}$ is ambiguous. One might try to fix the ambiguity by using again analytic continuation, but here we will content ourselves with the identification of beta functions and with the computation of the constant Love numbers only when these beta functions vanish.

\subsection{Einstein gravity}
Let us briefly review the computation of Love numbers in GR before considering the higher-derivative corrections.  We consider the static Teukolsky equation \req{Teuk2} with $\alpha=0$. The general solution is expressed in terms of the associated Legendre polynomials $P_{\ell}^{2}$ and associated Legendre functions of the second kind $Q_{\ell}^{2}$ as
\begin{equation}
\psi=\Delta\left[N_{1}  P_{\ell}^{2}\left(\frac{r}{M}-1\right)+N_{2}Q_{\ell}^{2}\left(\frac{r}{M}-1\right)\right]\, ,
\end{equation} 
where $N_{1}$ and $N_{2}$ are two integration constants. 
Now, the functions $Q_{\ell}^{2}(x)$ contain $\log(x-1)$ terms in their expansion near $x=1$, and therefore the solution with $N_{2}$ is singular at the horizon $r=2M$. Thus, we set $N_{2}=0$. On the other hand, we normalize the solution such that $\psi\sim(r/M)^{\ell+2}$ for $r\to\infty$.  This fixes the value of $N_{1}$ and we find
\begin{equation}\label{psih}
\psi^{(0)}=-\frac{2^{\ell}\ell!(\ell-2)!\Delta}{(2\ell)! M^2}  P_{\ell}^{2}\left(\frac{r}{M}-1\right)\, ,
\end{equation} 
where the superscript $^{(0)}$ denotes that this the GR solution, so it is zeroth-order in the higher-derivative couplings.  

For integer $\ell$, the Legendre functions $P_{\ell}^{2}$ are in fact polynomials, thus indicating that the Love numbers of Schwarzschild black holes vanish, $k_{\ell}=0$. However, as discussed above, this statement is coordinate dependent. To compute the Love numbers in a gauge-independent form, we promote $\ell$ to a real number $\hat{\ell}$ and perform the expansion around $r\to\infty$ analytically. We find
\begin{equation}\label{eq:asympt0}
\begin{aligned}
\psi^{(0)}(r)&=\left(\frac{r}{M}\right)^{\hat{\ell}+2}\left[1+\mathcal{O} \left(\frac{M}{r}\right)+\ldots\right]\\
&+\frac{\Gamma(-1/2-\hat{\ell})\Gamma(\hat{\ell}-1)}{2^{2\hat{l}+1}\Gamma(-\hat{\ell}-2)\Gamma(\hat{\ell}+1/2)} \left(\frac{M}{r}\right)^{\hat{\ell}-1}\left[1+\mathcal{O} \left(\frac{M}{r}\right)+\ldots\right]\, .
\end{aligned}
\end{equation}
Thus we identify the Love numbers
\begin{equation}
k_{\hat{\ell}}=\frac{\Gamma(-1/2-\hat{\ell})\Gamma(\hat{\ell}-1)}{2^{2\hat{\ell}+2}\Gamma(-\hat{\ell}-2)\Gamma(\hat{\ell}+1/2)}\, ,
\end{equation}
which in fact vanish when $\hat{\ell}=\ell$ is a positive integer, since $\Gamma(-\ell-2)=\infty$. 

\subsection{Computation of the corrections}
We now include the corrections to the Teukolsky equation in \req{Teuk2}. As we show in a moment, the corrections can be recast as a source term in the uncorrected Teukolsky equation. 
Thus, we start by introducing a basic result about the solutions of such equation. 

We introduce the notation
\begin{equation}
\mathcal{O}_{s}\psi\equiv  \Delta^{-s}\frac{d}{dr}\left(\Delta^{1+s}\frac{d\psi}{dr}\right)+V_{\ell}\psi\, ,
\end{equation}
denoting the uncorrected Teukolsky operator for spin $s$, 
and consider the static Teukolsky equation with a source $S(r)$
\begin{equation}
\mathcal{O}_{s}\psi=S(r)\, .
\end{equation}
As we show in Appendix \ref{app:sol}, the solution to this equation that is regular at the horizon and behaves at infinity as  $\psi\sim (r/M)^{\ell+2}$, is given by
\begin{equation}\label{gensol}
\psi=\psi^{(0)}+\psi_{p}[S]\, ,
\end{equation}
where
\begin{equation}\label{psip}
\psi_{p}[S]\equiv -\psi^{(0)}(r)\int_{r}^{\infty}\frac{dr''}{\Delta^{s+1}(r'')\left(\psi^{(0)}(r'')\right)^2}\int_{r_{+}}^{r''}dr' S(r')\psi^{(0)}(r')\Delta^{s}(r')\, .
\end{equation}
The subscript $p$ here denotes that this is a particular solution of the inhomogeneous Teukolsky equation. 

We can apply this result straightforwardly to obtain the solution of \req{Teuk2}. We are only interested in the effect of higher-derivative corrections at first order in the couplings so we write
\begin{equation}
\psi=\psi^{(0)}+\alpha\psi^{(1)}+\mathcal{O}(\alpha^2)\, .
\end{equation}
Expanding \req{Teuk2} at first order in $\alpha$, then we see that  $\psi^{(1)}$ satisfies the equation
\begin{equation}\label{eqpsi1}
\mathcal{O}_{s}\psi^{(1)}=-\delta V_{\ell} \psi^{(0)}\, .
\end{equation}
Therefore, the solution for $\psi^{(1)}$ is simply
\begin{equation}\label{psi1}
\psi^{(1)}=-\psi_{p}[\delta V_{\ell} \psi^{(0)}]\, .
\end{equation}
Thus, we just have to plug in $S=\delta V_{\ell} \psi^{(0)}$ with $\delta V_{\ell}$ given by \req{dVom0} and $\psi^{(0)}$ given by \req{psih} into \req{psip} and carry out the integration in order to obtain the solution. Now, the discussion is different depending on the value of the constant $C_{\ell}$ entering in \req{dVom0}.

\subsubsection{Case $C_\ell\neq 0$: running of Love numbers}
For integer values of $\ell$, the integration of \req{psi1} can be carried out analytically, and we show the explicit form of $\psi^{(1)}$ in Appendix \ref{app:sol} for a few values of $\ell$. 
The asymptotic expansion of these solutions contains a logarithmic term in the coefficient of $r^{1-\ell}$ whenever $C_{\ell}\neq 0$. For instance, for $\ell=2$ we have
 \begin{equation}\label{psi12}
 \begin{aligned}
\psi^{(1)}_{\ell=2}=&\frac{r^3 B_2}{4M^3}+\frac{r^2}{12M^2}\left(2 C_2-11 B_2\right)+\frac{r}{3M}\left(2 B_2-C_2\right)+\frac{1}{6} \left(2 B_2-7 C_2\right)\\
&+\frac{2 C_{2} M \left(47+60 \log (r/M)\right)}{75 r}+\mathcal{O}(r^{-2})\, .
 \end{aligned}
\end{equation}
According to our discussion at the beginning of the section, this means that the Love number $\ell=2$ runs. Using the definition in \req{betadef}, we identify the beta function $\beta_{2}=4 C_{2}/5$. Similarly, for the next few values of $\ell$ we find
\begin{equation}\label{betapart}
\beta_{2}=\frac{4}{5}C_{2}\, ,\quad \beta_{3}=\frac{4}{7}C_{3}\, ,\quad \beta_{4}=\frac{16}{49}C_{4}\, ,\quad \beta_{5}=\frac{16}{99}C_{5}\, , \quad \ldots
\end{equation}
In fact, after computing a dozen coefficients it is possible to guess the general pattern, which is given by
\begin{equation}\label{betagen}
\beta_{\ell}=\frac{4^{\ell-2}(\ell!)^4(\ell+1)^2(\ell+2)^2}{(1+2\ell)(2\ell)!^2}C_{\ell}\, .
\end{equation}
We note that this result is gauge invariant. To see this,  observe that in the asymptotic expansion of $\psi^{(1)}$, there are no logarithms in any of the coefficients of $r^{n}$ with $n>1-\ell$.  
This implies that the coefficient of $r^{1-\ell}\log(r/M)$ --- the beta function --- cannot be modified by coordinate transformations that are regular at infinity, like \req{radialchange}.  Therefore, the definition of $\beta_{\ell}$ is robust.
This result implies that Love numbers run whenever $C_{\ell}\neq 0$.

\subsubsection{Case $C_\ell=0$: Love numbers via analytic continuation}
Things get more interesting when $C_{\ell}=0$. If we look at the explicit solutions for $\psi^{(1)}$ in Appendix \ref{app:sol}, we can see that they become polynomials in $r$ when $C_{\ell}=0$. In particular, there is no $r^{1-\ell}$ term so it would seem that the Love number vanishes. However, we must take into account the ambiguities in the definition of Love numbers. In fact, we recall that transformations of the form \req{Psitrans} have been applied to the variable $\psi$ in order to set the modified Teukolsky equation in the form \req{Teuk2}. In addition,  the radial coordinate that we are using is not the usual Schwarzschild coordinate.  As a consequence, the coefficient of $r^{1-\ell}$ is a meaningless quantity.  
In order to extract the actual Love numbers, we resort to the magic of analytic continuation. Thus, we promote $\ell$ to take arbitrary real values $\ell\to \hat{\ell}$ and then we look for the coefficient of $r^{1-\hat{\ell}}$. As we discussed earlier, this provides a gauge-independent notion of Love numbers, independent of the variable $\psi$ or the radial coordinate employed. The main obstacle, however, is that this method requires that we have an analytic expression  for the asymptotic expansion of \req{psi1} for arbitrary $\hat{\ell}$. This is challenging, even for the relatively simple form of the source term. 

Here we propose a simple way to identify the Love numbers via analytic continuation without the need to obtain the solution for arbitrary $\hat{\ell}$.\footnote{Another interesting method has been recently proposed by \cite{Barbosa:2025uau}, making use of Green functions in the context of Regge-Wheeler-Zerilli equations.} The idea is to consider values of $\hat{\ell}$ arbitrarily close to a particular integer, 
\begin{equation}\label{leps}
\hat{\ell}=\ell+\epsilon\, ,
\end{equation} 
and expand perturbatively in $\epsilon$. Let us see how this works. We assume that the analytically continued solution takes the form
\begin{equation}\label{Psihatl}
\psi_{\hat{\ell}}=\left(\frac{r}{M}\right)^{\hat{\ell}+2}\left[1+\sum_{n=1}^{\infty}a_{n}(\hat \ell)\left(\frac{M}{r}\right)^{n}\right]+\left(\frac{r}{M}\right)^{1-\hat{\ell}}\left[b_{1}(\hat{\ell})+\sum_{n=1}^{\infty}b_{n}(\hat \ell)\left(\frac{M}{r}\right)^{n}\right]\, .
\end{equation}
According to our definition \req{LoveDef}, the Love number would be given by $k_{\hat \ell}=b_{1}(\hat \ell)/2$, and one would obtain the physical Love number by evaluating this expression on integer $\hat{\ell}=\ell$. Observe that, if one were to identify the Love number by looking at the solution for a particular integer $\ell$, one would obtain the wrong answer $2k_{\ell}=b_{1}(\ell)+a_{2\ell+1}(\ell)$, so analytic continuation is necessary. 
Let us then expand \req{Psihatl} around a specific $\ell$. At first order in $\epsilon$, the solution reads
\begin{equation}\label{Psihatl2}
\begin{aligned}
\psi_{\ell+\epsilon}&=\left(\frac{r}{M}\right)^{\ell+2}\left[1+\epsilon\log\left(\frac{r}{M}\right)\right]+\ldots\\
&+\left(\frac{r}{M}\right)^{1-\ell}\left[b_{1}(\ell)+a_{2\ell+1}(\ell)+\epsilon \left(\left(a_{2\ell+1}(\ell)-b_{1}(\ell)\right)\log\left(\frac{r}{M}\right)+\text{const}\right)\right]+\ldots+\mathcal{O}(\epsilon^2)\, .
\end{aligned}
\end{equation}
Therefore, we can determine the Love number from the coefficient of $r^{1-\ell}$ for $\epsilon=0$ and from the coefficient of $\epsilon r^{1-\ell}\log(r/M)$. Namely, if we get that our solution behaves as 

\begin{equation}\label{Psihatl3}
\begin{aligned}
\psi_{\ell+\epsilon}&=\left(\frac{r}{M}\right)^{\ell+2}\left[1+\epsilon\log\left(\frac{r}{M}\right)\right]+\ldots\\
&+\left(\frac{r}{M}\right)^{1-\ell}\left[p_{\ell}+\epsilon \left(q_{\ell}\log\left(\frac{r}{M}\right)+\text{const}\right)\right]+\ldots+\mathcal{O}(\epsilon^2)\, ,
\end{aligned}
\end{equation}
for certain coefficients $p_{\ell}$, $q_{\ell}$, then we read off the Love number as
\begin{equation}\label{kPQ}
k_{\ell}=\frac{p_{\ell}-q_{\ell}}{4}\, .
\end{equation}

Let us apply this idea to the Teukolsky equation in the cases in which $C_{\ell}=0$. First we analytically coninue $\ell\to\hat{\ell}$ in the equation \req{Teuk2}. Then we consider $\hat{\ell}$ around an integer \req{leps} and we expand the solution simultaneously in $\epsilon$ and in the higher-order coupling $\alpha$. Thus, we write
\begin{equation}
\psi_{\ell+\epsilon}=\psi^{(0)}+\alpha \psi^{(1)}+\epsilon\left(\psi_{1}^{(0)}+\alpha \psi_{1}^{(1)}\right)+\mathcal{O}(\alpha^2,\epsilon^2)\, ,
\end{equation}
where the subscript in $\psi_{1}^{(i)}$ denotes first order in $\epsilon$. 
Inserting this into \req{Teuk2} and expanding, leads to the equation \req{eqpsi1} and to the equations
\begin{align}\label{eq10}
\mathcal{O}_{-2}\psi_{1}^{(0)}=&-V_{\ell}' \psi^{(0)}\, ,\\
\mathcal{O}_{-2}\psi_{1}^{(1)}=&-\delta V_{\ell}\psi_{1}^{(0)}-\psi^{(0)}\delta V_{\ell}'-\psi^{(1)}V_{\ell}'\, .
\label{eq11}
\end{align}
Here, for any quantity $X_{\ell}$ that depends on $\ell$, $X_{\ell}'$ denotes a derivative of its analytically continued version,
\begin{equation}
X_{\ell}'=\frac{dX_{\hat{\ell}}}{d\hat{\ell}}\bigg|_{\hat{\ell}=\ell}\, . 
\end{equation}
Thus we have
\begin{equation}
V_{\ell}'=-1-2\ell\, ,\quad \delta V_{\ell}'=B_{\ell}'\frac{M}{ r}+C_{\ell}'\frac{M^2\left(r+2M\right)}{r^3}\, .
\end{equation}
Observe that for a given $\ell$ one may have $C_{\ell}=0$ but $C_{\ell}'\neq 0$. In fact, for the higher-derivative corrections we observe that $C_{2}=0$ for the cubic theories, and $C_{2}=C_{3}=0$ for the quartic ones. 

The equations \req{eq10} and \req{eq11} can again be solved straightforwardly applying \req{psip} and we show their exact solution for $\ell=2, 3 ,4$ in Appendix \ref{app:sol}. The asymptotic expansion of these solutions reads

\begin{align}\notag
\psi_{2+\epsilon}&=\left(\frac{r}{M}\right)^{4}\left[1+\epsilon\log\left(\frac{r}{M}\right)\right]+\ldots\\
&+\frac{2\alpha \epsilon\, C_{2}'}{75}\left(\frac{M}{r}\right)\left[47+60 \log\left(\frac{r}{M}\right)\right]+\ldots+\mathcal{O}(\alpha^2,\epsilon^2)\, ,\\\notag
\psi_{3+\epsilon}&=\left(\frac{r}{M}\right)^{5}\left[1+\epsilon\log\left(\frac{r}{M}\right)\right]+\ldots\\
&+\frac{2\alpha \epsilon\, C_{3}'}{735}\left(\frac{M}{r}\right)^2\left[-101+420 \log\left(\frac{r}{2M}\right)\right]+\ldots+\mathcal{O}(\alpha^2,\epsilon^2)\, \\\notag
\psi_{4+\epsilon}&=\left(\frac{r}{M}\right)^{6}\left[1+\epsilon\log\left(\frac{r}{M}\right)\right]+\ldots\\
&+\frac{4\alpha \epsilon\, C_{4}'}{15435}\left(\frac{M}{r}\right)^3\left[-1901+2520 \log\left(\frac{r}{2M}\right)\right]+\ldots+\mathcal{O}(\alpha^2,\epsilon^2)\, .
\end{align}
Therefore, comparing with \req{Psihatl3} we read off $p_{2}=p_{3}=p_{4}=0$, $q_{2}=8\alpha C_{2'}/5$, $q_{3}=8\alpha C_{3'}/7$, $q_{4}=32\alpha C_{4'}/49$, and consequently, according to \req{kPQ}, the Love numbers are 
\begin{equation}
k_{2}=-\frac{2}{5}\alpha C_{2}'\, ,\quad k_{3}=-\frac{2}{7}\alpha C_{3}'\, ,\quad k_{4}=-\frac{8}{49}\alpha C_{4}'\, .
\end{equation}
This pattern is the same one as with the beta functions \req{betapart} but with $C_{\ell}\to -C_{\ell}'/2$. This is not a coincidence: the part of the source term proportional to $C_{\ell}'$ in \req{eq11} is the same as the part of the source term proportional to $C_{\ell}$ in \req{eqpsi1}. Therefore, similarly to \req{betagen}, we obtain the general expression
\begin{equation}\label{lovegen}
k_{\ell}=-\frac{4^{\ell-2}(\ell!)^4(\ell+1)^2(\ell+2)^2}{2(1+2\ell)(2\ell)!^2}C_{\ell}'\, ,\quad \text{whenever}\,\,\, C_{\ell}=0\, .
\end{equation}

\section{Love numbers in the EFT of GR}\label{sec:loveEFT}
We recap the results of the last two sections. The static perturbations of non-rotating black holes in the EFT extension of GR are ruled by the master equation \req{Teuk2}, that depends on certain coefficients $B_{\ell}$ and $C_{\ell}$. 
We have computed the Love numbers for general values of these coefficients, and we have found that
\begin{itemize}
\item When $C_{\ell}\neq 0$ the Love numbers run and their beta functions are given by \req{betagen}. 
\item When $C_{\ell}=0$ the Love numbers are constant and given by \req{lovegen},  that depends on $C_{\ell}'=dC_{\ell}/d\ell$. 
\end{itemize}
We can now use the specific values of $C_{\ell}$ for each higher-derivative correction --- we show them in Appendix \ref{app:coef} ---  in order to obtain the explicit values of the TLNs. However, in order to obtain the correct interpretation of \req{lovegen} we first have to take into account the parity content of the gravitational perturbations described by the Teukolsky equation. 

\subsection{Parity considerations: three types of Love numbers}\label{sec:polarization}
The TLNs \req{lovegen} obtained from the Teukolsky equation, or the corresponding beta functions \req{betagen}, actually describe different types of tidal deformations depending on the polarization of the perturbation. To see this, we must take a look at the form of metric perturbations. 
Schematically, we can imagine that our metric perturbation is decomposed as the sum of a tidal field plus a response field, 
\begin{equation}
h_{\mu\nu}=h_{\mu\nu}^{T}+h_{\mu\nu}^{R}\, ,
\end{equation}
and each of these is decomposed in an even-parity (polar-type) part $h_{\mu\nu}^{+}$ and an odd-parity (axial-type) part $h_{\mu\nu}^{-}$, 
\begin{equation}
h_{\mu\nu}^{T}=a_{+}h_{\mu\nu}^{T+}+a_{-}h_{\mu\nu}^{T-}\, ,\quad h_{\mu\nu}^{R}=b_{+}h_{\mu\nu}^{R+}+b_{-}h_{\mu\nu}^{R-}\, .
\end{equation}
Here we assume that the components $h_{\mu\nu}^{T\pm}$, $h_{\mu\nu}^{R\pm}$ are normalized in some appropriate way. 
The coefficients $a_{\pm}$ are free and determine the form of the tidal field, while the coefficients $b_{\pm}$ must be a linear function of the former. Therefore, there is a tidal matrix $\mathbb{T}$  such that
\begin{equation}
\begin{pmatrix}
b_{+}\\
b_{-}
\end{pmatrix}=\mathbb{T}
\begin{pmatrix}
a_{+}\\
a_{-}
\end{pmatrix}\, .
\end{equation}
Assuming that $\mathbb{T}$ is symmetric (we will see this is the case for higher-derivative corrections), we denote the components of the tidal matrix by
\begin{equation}\label{transfer}
\mathbb{T}=
\begin{pmatrix}
k^{+}& k^{\rm mix}\\
k^{\rm mix}& k^{-}
\end{pmatrix}\, .
\end{equation}
Here $k^{+}$ represents the polar response due to a polar tide, and it is therefore an  ``electric-type'' Love number, while $k^{-}$ represents the axial response due to an axial tide, and it is identified with a  ``magnetic-type'' Love number.  The non-diagonal component $k^{\rm mix}$ is a bit more exotic as it means that a polar tidal field induces an axial response and vice-versa. 

The result \req{lovegen} from the Teukolsky equation  captures the three types of Love numbers $k^{+}$, $k^{-}$ and $k^{\rm mix}$. In order to see to which type of TLN \req{lovegen} corresponds to, we have to connect the Teukolsky variable with metric perturbations. \\

At the level of the Teukolsky equation, the polarization of the perturbation is encoded in a parameter $q_{-2}$ that appears in the expressions of $C_{\ell}$ and $B_{\ell}$. For a detailed definition of this parameter we refer to section 4.2 of \cite{Cano:2023tmv}, but give here an intuitive explanation.  In GR (without including the corrections yet), the curvature perturbations that we are studying derive from a metric perturbation of the form
\begin{equation}
h_{\mu\nu}=\mathcal{O}_{\mu\nu}\psi_{-2}+\bar{\mathcal{O}}_{\mu\nu}\psi_{-2}^{*}\, ,
\end{equation}
where $\mathcal{O}_{\mu\nu}$ is a certain operator whose form is not relevant for our discussion, $\bar{\mathcal{O}}_{\mu\nu}$ is its complex conjugate and $\psi_{-2}$, $\psi_{-2}^{*}$ are Hertz potentials. Both $\psi_{-2}$ and $\psi_{-2}^{*}$ satisfy the Teukolsky equation of spin $s=-2$, and they are proportional to the Teukolsky variable $\Psi_{4}$ and to each other.\footnote{More precisely, we are referring to the radial part of each variable.} The parameter $q_{-2}$ is precisely the proportionality constant, $\psi_{-2}^{*}=q_{-2}\psi_{-2}$. Therefore, the metric perturbation takes the form 
\begin{equation}\label{hmunuq2}
h_{\mu\nu}=\mathcal{O}_{\mu\nu}\psi_{-2}+q_{-2}\bar{\mathcal{O}}_{\mu\nu}\psi_{-2}\, .
\end{equation}
As observed in \cite{Cano:2023tmv}, the choices $q_{-2}=\pm1$ correspond to perturbations of defined parity: $q_{-2}=1$ corresponds to polar perturbations and $q_{-2}=-1$ to axial ones. Therefore, the real and imaginary parts of $\mathcal{O}_{\mu\nu}\psi_{-2}$ are neatly identified with the even and odd parity contents of the metric perturbation, 
\begin{equation}
\mathcal{O}_{\mu\nu}\psi_{-2}=\frac{1}{2}\left(h_{\mu\nu}^{+}+i h_{\mu\nu}^{-}\right)\, ,\quad \bar{\mathcal{O}}_{\mu\nu}\psi_{-2}=\frac{1}{2}\left(h_{\mu\nu}^{+}-i h_{\mu\nu}^{-}\right)\, .
\end{equation}
Then, we can write the general form of the metric perturbation \req{hmunuq2} as
\begin{equation}\label{hmunuq22}
h_{\mu\nu}=\frac{1}{2}(1+q_{-2})h_{\mu\nu}^{+}+\frac{i}{2}(1-q_{-2})h_{\mu\nu}^{-}\, .
\end{equation}
Thus, the parameter $q_{-2}$ determines the weight of each perturbation type in the solution. 

Now there is a crucial observation. In GR, the value of $q_{-2}$ is arbitrary due to isospectrality (the Teukolsky equation is independent of $q_{-2}$), but in extensions of GR, the value of $q_{-2}$ is fixed when we look for eigenmodes (in our case, tidal modes). The argument is exactly the same as in the case of quasinormal modes, which has been recently discussed in Refs.~\cite{Cano:2023tmv,Cano:2023jbk,Cano:2024ezp}.  
The idea is that the Teukolsky equation for $\Psi_{4}$ and that for the conjugate variable $\Psi_{4}^{*}$ must be consistent with each other, since they describe the same gravitational perturbation. In our case this means that they must predict the same Love numbers (and in the case of quasinormal modes it means that both must share the same spectrum).  We saw in section \ref{sec:modifiedTeuk} that the modified Teukolsky equation for $\Psi_{4}^{*}$ takes the same form as the one for $\Psi_{4}$ but with different coefficients $B_{\ell}^{*}$, $C_{\ell}^{*}$ given by \req{coefdual}. Since the Love numbers only depend on $C_{\ell}$, the consistency condition that both equations yield the same results is
\begin{equation}\label{cons}
\bar{C}_{\ell}(1/q_{-2})=C_{\ell}(q_{-2})\, .
\end{equation}
Moreover, this automatically implies $\bar{B}_{\ell}(1/q_{-2})=B_{\ell}(q_{-2})$ so the equations of $\Psi_{4}$ and $\Psi_{4}^{*}$ become identical whenever \req{cons} is satisfied. 
The consistency condition fixes the allowed values of the $q_{-2}$ parameter and therefore determines the polarization content of the eigenmodes.  
The discussion is different depending on the type of higher-derivative corrections. 

In the case of higher-derivative corrections that preserve parity, the two possible solutions of \req{cons} are $q_{-2}=\pm 1$ (it is immediate to confirm this by looking at the expressions for $C_{\ell}$ in Appendix \ref{app:coef}), so that each tidal mode has a definite parity. This is the usual situation in which an even-parity tidal field generates an even-parity response and the same with axial perturbations. Therefore, the tidal matrix is diagonal and we simply identify the electric and magnetic Love numbers as
\begin{equation}\label{kpm}
k^{+}_{\ell}=k_{\ell}\big|_{q_{-2}=+1}\, ,\quad k^{-}_{\ell}=k_{\ell}\big|_{q_{-2}=-1}
\end{equation}

For theories that break parity, things are different. We can see from the expressions $C_{\ell}$ in the appendix that $C_{\ell}\propto i/q_{-2}$ in those theories. Therefore, the solutions of the consistency condition \req{cons} are $q_{-2}=\pm i$. Plugging these values in \req{hmunuq22}, we obtain the two tidal modes 
\begin{equation}\label{hmunuq23}
h_{\mu\nu}^{(1)}\propto h_{\mu\nu}^{+}+h_{\mu\nu}^{-}\, ,\quad h_{\mu\nu}^{(2)}\propto h_{\mu\nu}^{+}-h_{\mu\nu}^{-}\, ,
\end{equation}
that in this case consist of a combination of polar and axial perturbations. Furthermore, since $C_{\ell}\propto i/q_{-2}$, these modes have opposite Love numbers. Thus, the solutions behave schematically as
\begin{equation}
h_{\mu\nu}^{(1)}=h_{\mu\nu}^{T+}+h_{\mu\nu}^{T-}+k_{\ell} \left(h_{\mu\nu}^{R+}+h_{\mu\nu}^{R-}\right)\, ,\quad h_{\mu\nu}^{(2)}=h_{\mu\nu}^{T+}-h_{\mu\nu}^{T-}-k_{\ell} \left(h_{\mu\nu}^{R+}-h_{\mu\nu}^{R-}\right)\, .
\end{equation}
These are precisely the eigenmodes of a tidal matrix \req{transfer} with $k^{+}=k^{-}=0$ and
\begin{equation}\label{kmix}
k_{\ell}^{\rm mix}=k_{\ell}\big|_{q_{-2}=+i}\, .
\end{equation}
Therefore, in a theory with parity-breaking corrections, axial tidal fields induce a polar response, and vice-versa.  
\subsection{Results}
We can now combine the expressions \req{lovegen}, \req{betagen} with the expressions of $C_{\ell}$ in Appendix \ref{app:coef} in order to generate the different Love numbers and beta functions. For parity-preserving corrections, the electric and magnetic Love numbers are given by \req{kpm}, while for parity-breaking corrections the ``mixing'' Love numbers are given by \req{kmix}.

We observe that $C_{2}=0$ for cubic corrections while $C_{2}=C_{3}=0$ for quartic corrections. All the other values of $C_{\ell}$ are non-vanishing.  Therefore, Love numbers with $\ell\ge 3$ run in cubic gravity and those with $\ell\ge 4$ run in quartic gravity. All the non-running TLNs are collected in Table \ref{table:TLNall}. 

  \begingroup
 \setlength{\tabcolsep}{8pt} 
 \renewcommand{\arraystretch}{1.8} 
 \begin{table}[t!]
 	\centering
 		\begin{tabular}{|c"c|c"c|c"c|c|}
 			\hline
 			Theory & $k_{2}^{+}$ & $k_{3}^{+}$& $k_{2}^{-}$ & $k_{3}^{-}$& $k_{2}^{\rm mix}$ & $k_{3}^{\rm mix}$  \\ \hline\hline
 			$\lambda_{\rm ev}$ & $\displaystyle 28\frac{\lambda_{\rm ev}}{M^4}$ & runs & $\displaystyle-20\frac{\lambda_{\rm ev}}{M^4}$& runs & 0& 0  \\ 
			$\lambda_{\rm odd}$ & $0$ & $0$ & $0$ & $0$ & $\displaystyle -24\frac{\lambda_{\rm odd}}{M^4}$& runs \\ 
 			$\xi_{1}$ &  $\displaystyle\frac{1008}{25}\frac{\xi_{\rm 1}}{M^6}$ & $\displaystyle 1584\frac{\xi_{\rm 1}}{M^6}$ & $\displaystyle-\frac{432}{25}\frac{\xi_{\rm 1}}{M^6}$& $\displaystyle -\frac{6672}{7}\frac{\xi_{\rm 1}}{M^6}$& 0& 0  \\ 
 			$\xi_{2}$ &  $0$ & $0$ & $\displaystyle\frac{96}{5}\frac{\xi_{\rm 2}}{M^6}$& $\displaystyle\frac{4128}{7}\frac{\xi_{\rm 2}}{M^6}$ & 0& 0  \\ 
			$\xi_{3}$ &  $0$ & $0$ & $0$& $0$ &$-\displaystyle\frac{96}{5}\frac{\xi_{\rm 3}}{M^6}$ & $-\displaystyle\frac{5472}{7}\frac{\xi_{\rm 3}}{M^6}$  \\ \hline
			 	\end{tabular}
 	\caption{The analytical values of all the tidal Love numbers with $\ell=2,3$ in the EFT of GR.} 
 	\label{table:TLNall}
 \end{table}
  \endgroup 

Regarding the beta functions, we obtain the following general expressions in the case of cubic gravity 
\begin{align}
\beta^{+}_{\ell}&= -\frac{7\lambda_{\rm ev}}{6M^4} F_{\ell} \left(\ell(\ell+1)-4\right) \,, \\\label{betameven}
\beta^{-}_{\ell}&= \frac{5\lambda_{\rm ev}}{6M^4} F_{\ell} \left(\ell(\ell+1)-4\right) \, ,\\
\beta^{\rm mix}_{\ell}&=\frac{\lambda_{\rm odd}}{M^4} F_{\ell} \left(\ell(\ell+1)-4\right)\, ,
\end{align}
where $F_{\ell}$ is the coefficient 
\begin{equation}
F_{\ell}=\frac{4^{\ell-3} (\ell!)^4  (\ell-2) (\ell-1) \ell (\ell+1)^3 (\ell+2)^3 (\ell+3) }{(2 \ell+1) (2 \ell)!^2}\, .
\end{equation}
For the quartic theories, we get
\begin{align}
\beta^{+}_{\ell}&=\frac{(\ell-3)(\ell+4)F_{\ell}}{16} \frac{\xi_{1}}{M^6}\left(\frac{8}{3}-\frac{106 L}{15}+\frac{1771 L^2}{675}-\frac{143 L^3}{675}\right) \,, \\\notag
\beta^{-}_{\ell}&= \frac{(\ell-3)(\ell+4)F_{\ell}}{16} \left[\frac{\xi_{1}}{M^6}\left(-\frac{136}{9}+\frac{17198 L}{1575}-\frac{35477 L^2}{14175}+\frac{2353 L^3}{14175}\right)\right.\\
&\left.-\frac{8 (L-2)L (13 L-113)}{1575}\frac{\xi_{2}}{M^6}\right] \, ,\\
\beta^{\rm mix}_{\ell}&=\frac{(\ell-3)(\ell+4)F_{\ell}}{16} \frac{\xi_{3}}{M^6}\left(-\frac{40}{9}+\frac{7534 L}{1575}-\frac{20669 L^2}{14175}+\frac{1573 L^3}{14175}\right)\, ,
\end{align}
where $L=\ell(\ell+1)$.

For convenience, we give the first non-vanishing values of the beta functions. For cubic gravity, these are the ones with $\ell=3$,  
\begin{align}
\beta^{+}_{3}&= -960\frac{\lambda_{\rm ev}}{M^4} \,, \quad
\beta^{-}_{3}= \frac{4800}{7}\frac{\lambda_{\rm ev}}{M^4}\, ,\quad
\beta^{\rm mix}_{3}=\frac{5760}{7}\frac{\lambda_{\rm odd}}{M^4}\, ,
\end{align}
while for quartic corrections, the first non-vanishing ones are those with $\ell=4$
\begin{align}
\beta^{+}_{4}&=-161280\frac{\xi_{\rm 1}}{M^6}\, ,\quad
\beta^{-}_{4}=109056\frac{\xi_{\rm 1}}{M^6}-55296\frac{\xi_{\rm 2}}{M^6} \, ,\quad
\beta^{\rm mix}_{4}=81408\frac{\xi_{\rm 3}}{M^6}\, .
\end{align}

Some of these TLNs and beta functions have been computed through the Regge-Wheeler-Zerilli approach in previous literature.  Let us compare our results with those.  
For the parity-preserving quartic theories, our results for $k_{2}^{\pm}$ agree with those of \cite{Cardoso:2018ptl} if we take into account that our couplings $\xi_{1}$, $\xi_{2}$ are related to theirs ($\epsilon_{1}$, $\epsilon_{2}$)  by $\xi_{1}M^{-6}=-\epsilon_{1}$, $\xi_{2}M^{-6}=-4\epsilon_{2}$. Refs.~\cite{Katagiri:2023umb,Katagiri:2024fpn} additionally obtained the values of  $k_{3}^{\pm}$, and our results in Table~\ref{table:TLNall} also agree with them if we take into account that, besides the different definition of the coupling constants,  their conventions for Love numbers include a factor of $2^{-2\ell-1}$ relative to ours. Our value for the beta function $\beta^{-}_{4}$ for the theory $\xi_{2}$ is also consistent with the numeric value found in \cite{Katagiri:2023umb}, taking into account the same differences in conventions. 

Regarding cubic gravities, our results for $k_{2}^{\pm}$ for $\lambda_{\rm ev}$ manifestly agree with the values found in Ref.~\cite{Cai:2019npx}. Refs.~\cite{DeLuca:2022tkm,Barbosa:2025uau} computed $k_{2}^{-}$ for the same theory and \cite{Barbosa:2025uau} obtained a general formula for $\beta^{-}_{\ell}$, but they use different conventions. To perform the comparison, we take into account that these references define their Love numbers --- let us denote them here $\hat{k}_{\ell}$ --- by the following expansion of the Regge-Wheeler variable
$\psi^{\rm RW}\sim (r/r_{+})^{\ell+1}+\hat{k}_{\ell}^{-}(r/r_{+})^{-\ell} +\ldots$, 
with $r_{+}=2M$.
On the other hand, in the way we are defining the Love numbers, they show up in the Regge-Wheeler variable as\footnote{For $\omega\to 0$, the relationship between the Regge-Wheeler variable and the radial Teukolsky function of $\Psi_{4}$  is $\psi\propto \frac{f(r)}{8}\left(\psi^{\rm RW} (6 M-\ell (\ell+1) r)+2 r (3 M-r) \frac{d\psi^{\rm RW}}{dr}\right)$} $\psi^{\rm RW}\sim (r/M)^{\ell+1}+2k^{-}_{\ell}\frac{(\ell+1)(\ell+2)}{\ell(\ell-1)}(r/M)^{-\ell}$. Therefore, both definitions are related by 
\begin{equation}
\hat{k}_{\ell}^{-}=2^{-2\ell}\frac{(\ell+1)(\ell+2)}{\ell(\ell-1)}k_{l}^{-}\, .
\end{equation}
In the case of the beta functions, there is an additional relative sign in the definition. 
Finally, we take into account that their coupling constant $\alpha$ is related to ours by $2\kappa^2\alpha=\lambda_{\rm ev}$, and we get an exact match. 

To the best of our knowledge, the Love numbers for parity breaking corrections  --- which are of a different type $k^{\rm mix}_{\ell}$ --- and the general  expressions for all the beta functions (except $\beta_{\ell}^{-}$ for $\lambda_{\rm ev}$) are shown here for the first time.

\section{Conclusions}\label{sec:conclusions}
We have shown how to identify the TLNs of non-rotating black holes in higher-derivative extensions of GR via the modified Teukolsky equation. This approach turns out to be very powerful, as it captures all types of higher-derivative corrections and all types of perturbations (axial/polar) in a single equation \req{Teuk2}. We have found that the Love numbers only depend on a coefficient $C_{\ell}$ entering in that equation: they run whenever $C_{\ell}\neq 0$ --- the beta functions are given in \req{betagen} --- and they have a constant value when $C_{\ell}=0$ --- see \req{lovegen}. In the latter case, it is crucial to use analytic continuation in $\ell$ in order to obtain a physically relevant (gauge invariant) result. To this end, we implemented a method consisting in expanding the solution around integer values of $\ell$, bypassing the difficulty of obtaining an analytic solution of the modified Teukolsky equation for arbitrary $\ell$. Combining our general formulas \req{betagen}, \req{lovegen}, with the expressions of $C_{\ell}$ predicted by higher-derivative corrections, we have obtained complete results for the TLNs in the EFT of GR.  These include non-vanishing values for the electric and magnetic-type TLNs, as a well as a new type of TLNs, denoted $k^{\rm mix}$, that arise in parity-violating theories. These have the effect of generating an axial response to a polar tidal field, and viceversa. 

We have compared our results with previous computations of Love numbers obtained via the Regge-Wheeler-Zerilli approach \cite{Cardoso:2018ptl,Cai:2019npx,DeLuca:2022tkm,Katagiri:2023umb,Katagiri:2024fpn,Barbosa:2025uau}, finding perfect agreement.
The matching of results is remarkable taking into account the we are using a completely different equation, expressed in different coordinates. This shows the power of analytic continuation in $\ell$ in order to obtain a gauge-invariant result, and gives robustness to all the results. Therefore, we expect that the Love numbers and beta functions that we have computed can be directly identified with the tidal coefficients of the worldline EFT \cite{Hui:2020xxx}. As a future goal, it would be interesting to confirm this result by providing an explicit match between black hole perturbation theory and the worldline EFT, although this is challenging.

We close by commenting on future directions. 
A relatively straightforward generalization of our analysis would allow one to compute the dissipative part of the black hole response \cite{Chia:2020yla,Goldberger:2020fot,Charalambous:2021mea}. To this end, one just needs to include the frequency dependence in the modified Teukolsky equation \req{Teuk} by using the $A_{k,\ell}(\omega)$ coefficients \cite{gitbeyondkerr} expanded to first order in $\omega$. The tidal dissipation constants $\nu_{\ell}$ could then be identified from the frequency-dependent Love numbers $k_{\rm \ell}(\omega)=k_{\ell}+i\nu_{\ell} M\omega+\mathcal{O}(\omega^2)$.  

More importantly, our results here set the stage for the computation of Love numbers of rotating black holes in the EFT of GR. The correction to the Teukolsky equation for rotating black holes takes in fact the same form as \req{eq:deltaV}, where the coefficients $A_{k,\ell}$ additionally depend on the angular momentum and on the harmonic number $m$. 
For the computation of beta functions, as we have seen, we only need to know these coefficients for a given $\ell$, so using the results of \cite{Cano:2024ezp}, available in \cite{gitbeyondkerr}, it should be moderately straightforward to obtain the beta functions for rotating black holes. The computation of the non-running Love numbers is considerably more involved. As our analysis has revealed, we need to know the analytic dependence of the modified Teukolsky equation on the angular number $\ell$ in order to identify $k_{\ell}$ --- see \req{lovegen}.   
Obtaining such analytic expression including the effect of rotation is computationally challenging, even if we restrict to a power expansion in the angular momentum. However, we do not foresee any fundamental obstacle other than the shear complexity of the calculations. We expect to report on this in the future.    

\section*{Acknowledgements}

I would like to thank Vitor Cardoso, Sylvain Fichet and Takuya Katagiri for useful correspondence and Marina David for feedback on an earlier version of this manuscript. 
The work of PAC received the support of a fellowship from “la Caixa” Foundation (ID 100010434) with code LCF/BQ/PI23/11970032. 

\appendix

\section{Coefficients of the potential}\label{app:coef}
We introduce
\begin{equation}
L= \ell(\ell+1)\, .
\end{equation}
For each of the higher-derivative Lagrangians in the EFT \req{eq:EFT}, the coefficients $B_{\ell}$, $C_{\ell}$ entering in the modified Teukolsky equation \req{Teuk2} are given by
\begin{align}
B_{\ell}^{\rm ev}&=\frac{\lambda_{\rm ev}}{M^4}\frac{(\ell-1) \ell (\ell+1) (\ell+2) \left(-60+418 L-201 L^2+21 L^3\right) \left(6+q_{-2}\right)}{840
   \left(L-1\right) q_{-2}}\, , \\
C_{\ell}^{\rm ev}&=-\frac{\lambda_{\rm ev}}{M^4}(\ell-2) (\ell-1) \ell (\ell+1) (\ell+2) (\ell+3) \left(L-4\right) \frac{\left(6+q_{-2}\right)}{24 q_{-2}}\, ,
\end{align}

\begin{align}
B_{\ell}^{\rm odd}&=-\frac{\lambda_{\rm odd}}{M^4}\frac{i (\ell-1) \ell (\ell+1) (\ell+2) \left(-60+418 L-201 L^2+21 L^3\right)}{140 \left(L-1\right) q_{-2}}\, , \\
C_{\ell}^{\rm odd}&=\frac{\lambda_{\rm odd}}{M^4}\frac{i (\ell-2) (\ell-1) \ell (\ell+1) (\ell+2) (\ell+3) \left(L-4\right)}{4 q_{-2}}\, ,
\end{align}

\begin{align}\notag
B_{\ell}^{\rm 1}&=\frac{\xi_{\rm 1}}{M^6}\frac{(\ell-1) \ell (\ell+1) (\ell+2)}{L-1} \Bigg[-\frac{73}{55}+\frac{18563 L}{3960}-\frac{1334789 L^2}{606375}+\frac{4303307
   L^3}{12474000}+\frac{182857 L^4}{31752000}\\\notag
   &-\frac{22987 L^5}{5292000}+\frac{13
   L^6}{60480}+\frac{1}{q_{-2}}\Bigg(-\frac{92}{55}-\frac{28753 L}{13860}+\frac{7609513 L^2}{1212750}-\frac{3824351
   L^3}{1247400}+\frac{9855127 L^4}{15876000}\\
   &-\frac{145933 L^5}{2646000}+\frac{1339
   L^6}{756000}\Bigg)\Bigg]\, , \\\notag
C_{\ell}^{\rm 1}&=-\frac{\xi_{\rm 1}}{M^6}(\ell-3) (\ell-2) (\ell-1) \ell (\ell+1) (\ell+2) (\ell+3) (\ell+4) \Bigg[\frac{7}{72}-\frac{1517 L}{50400}-\frac{857
   L^2}{907200}+\frac{13 L^3}{36288}\\\label{Cquartic1}
   &+\frac{1}{q_{-2}}\left(-\frac{5}{36}+\frac{3541 L}{25200}-\frac{18167
   L^2}{453600}+\frac{1339 L^3}{453600}\right)\Bigg]\, ,
\end{align}

\begin{align}\notag
B_{\ell}^{\rm 2}&=\frac{\xi_{\rm 2}}{M^6}\frac{(\ell-1) \ell (\ell+1) (\ell+2) \left(q_{-2}-1\right)}{\left(L-1\right) q_{-2}} \bigg(-\frac{62}{55}+\frac{167 L}{110}+\frac{317593 L^2}{1212750}-\frac{593
   L^3}{1980}\\
   &+\frac{73639 L^4}{882000}-\frac{643 L^5}{73500}+\frac{13 L^6}{42000}\bigg)
  \, , \\\label{Cquartic2}
C_{\ell}^{\rm 2}&=-\frac{\xi_{\rm 2}}{M^6}\frac{(\ell-3) (\ell-2) (\ell-1)^2 \ell^2 (\ell+1)^2 (\ell+2)^2 (\ell+3) (\ell+4) (13L-113) \left(q_{-2}-1\right)}{25200
   q_{-2}}\, ,
\end{align}

\begin{align}\notag
B_{\ell}^{\rm 3}&=\frac{\xi_{\rm 3}}{M^6}\frac{i (\ell-1) \ell (\ell+1) (\ell+2)}{\left(L-1\right) q_{-2}} \bigg(\frac{7}{5}+\frac{701 L}{2520}-\frac{360323 L^2}{110250}+\frac{381631
   L^3}{226800}-\frac{11180629 L^4}{31752000}\\
   &+\frac{169081 L^5}{5292000}-\frac{1573
   L^6}{1512000}\bigg)\, ,\\\notag
C_{\ell}^{\rm 3}&=\frac{\xi_{\rm 3}}{M^6}\frac{i (\ell-3) (\ell-2) (\ell-1) \ell (\ell+1) (\ell+2) (\ell+3) (\ell+4)}{q_{-2}} \bigg(-\frac{5}{72}+\frac{3767 L}{50400}\\\label{Cquartic3}
&-\frac{20669
   L^2}{907200}+\frac{1573 L^3}{907200}\bigg)\, .
\end{align}

\section{Solution of the perturbations}\label{app:sol}

\subsection{Solution of the inhomogeneous Teukolsky equation}
We consider the static ($\omega=0$) Teukolsky equation with a source term, 
\begin{equation}\label{Teukinhomo}
\Delta^{-s}\frac{d}{dr}\left(\Delta^{1+s}\frac{d\psi}{dr}\right)+V_{\ell}\psi = S(r)\, . 
\end{equation}
Here we consider perturbations of spin $s$ for generality, and $V_{\ell}=s(s+1)-\ell(\ell+1)$. Let us denote by $\psi^{(0)}$ the solution of the homogeneous equation that is regular at the horizon and normalized at infinity as $\psi^{(0)}\sim (r/M)^{\ell-s}$. Explicitly, the solution is
\begin{equation}
\psi^{(0)}=\frac{(i M)^s}{\Delta^{s/2}}\frac{2^{\ell}\ell!(\ell-s)!}{(2\ell)!} P_{\ell}^{s}\left(\frac{r}{M}-1\right)\, ,
\end{equation}
where $P_{\ell}^{s}$ are associated Legendre functions. Observe that for $r\to 2M$ this solution behaves as $\psi^{(0)}\sim (r-2M)^{-s}$ when $s<0$ and as $\psi^{(0)}\sim \text{const}+\mathcal{O}(r-2M)$ when $s\ge 0$, and it is smooth at the horizon.  
In order to obtain a solution of the inhomogeneous equation, we propose an ansatz of the form
\begin{equation}
\psi=\psi^{(0)} H\, .
\end{equation}
The equation \req{Teukinhomo} then becomes
\begin{equation}
\frac{1}{\psi^{(0)}\Delta^{s}}\frac{d}{dr}\left(\Delta^{1+s}\left(\psi^{(0)}\right)^2\frac{dH}{dr}\right)= S(r)\, ,
\end{equation}
which can be integrated straightforwardly. The first integral gives
\begin{equation}\label{eq:dH}
\frac{dH}{dr}=\frac{1}{\Delta^{1+s}(r)\left(\psi^{(0)}(r)\right)^2}\int_{r_{1}}^{r}dr' S(r')\Delta^s(r')\psi^{(0)}(r')\, ,
\end{equation}
where the limit of integration $r_{1}$ is an integration constant.  We want the solution to be smooth at $r=r_{+}=2M$, and since the term in front of the integral always diverges in that limit, the only way of to achieve a regular solution (assuming the source term is regular) is by making the integral vanish at $r=r_{+}$. Therefore, we must set $r_{1}=r_{+}$.  Integrating \req{eq:dH} then yields
\begin{equation}\label{eq:dH}
H=H_0+\int_{r_{2}}^{r}\frac{dr''}{\Delta^{1+s}(r'')\left(\psi^{(0)}(r'')\right)^2}\int_{r_{+}}^{r''}dr' S(r')\Delta^s(r')\psi^{(0)}(r')\, ,
\end{equation}
where for convenience we have introduced two integration constants $r_{2}$ and $H_0$, although of course they are equivalent. Since we want our solution to behave asymptotically as $\psi^{(0)}$ (as long as the source term decays fast enough at infinity), then we demand $H(r)\to 1$ when $r\to\infty$. Therefore, we set $H_{0}=1$ and $r_{2}=\infty$. This yields the result \req{gensol}.

\subsection{Explicit solutions}

The solutions of \req{psi1} with $\ell=2,3,4$ are given by
\begin{align}\notag
\psi^{(1)}_{\ell=2}&=B_{2}(x-1)^2\left(\frac{1}{3}+2 x\right)+C_{2}\Bigg[24 (x-1)^2 x^2p(x)+\left(-2-8 x+36 x^2-24 x^3\right)\log(x)\\\label{psi12}
&+(x-1)\left(\frac{10}{3}+\frac{56 x}{3}-24 x^2\right)\Bigg]\, ,\\\notag
\psi^{(1)}_{\ell=3}&=B_{3}(x-1)^2\left(-\frac{1}{15}-\frac{4 x}{5}+\frac{8 x^2}{3}\right)+C_{3}\Bigg[120 (x-1)^2 x^2 (2 x-1)p(x)\\\notag
&+\left(2+20 x-260 x^2+480 x^3-240 x^4\right)\log(x)\\\label{psi13}
&+(x-1)\left(-\frac{86}{15}-\frac{856 x}{15}+\frac{1504 x^2}{5}-240 x^3\right)\Bigg]\, ,\\\notag
\psi^{(1)}_{\ell=4}&=B_{4}(x-1)^2\left(\frac{2}{105}+\frac{8 x}{21}-\frac{22 x^2}{7}+4 x^3\right)+C_{4}\Bigg[720 (x-1)^2 x^2 \left(\frac{3}{7}+2 (x-1) x\right)p(x)\\\notag
&+\left(-\frac{12}{7}-\frac{216 x}{7}+\frac{5760 x^2}{7}-\frac{20640 x^3}{7}+3600 x^4-1440 x^5\right)\log(x)\\
&+(x-1)\left(\frac{212}{35}+\frac{504 x}{5}-\frac{8328 x^2}{7}+\frac{17648 x^3}{7}-1440 x^4\right)\Bigg]\, ,\label{psi14}
\end{align}
where $x=r/(2M)$ and
\begin{equation}
p(x)=\operatorname{Li}_{2}(1/x)-\log(x)\log(1-1/x)\, ,
\end{equation}
where $\operatorname{Li}_{2}$ is a polylogarithm. We note that the function $p(x)$ is smooth at the horizon $x=1$, and by extension all $\psi^{(1)}_{\ell}$ are. In fact these functions behave as $\psi^{(1)}_{\ell}\sim (x-1)^2$ near $x=1$, and they decay at infinity as $x^{l+1}$, so they do not change the asymptotic behavior of $\psi^{(0)}_{\ell}\sim x^{l+2}$.

The solutions of \req{eq10} read
\begin{align}
\psi^{(0)}_{1, \ell=2}&=\frac{4}{3} (x-1)^2 \left(-1-6 x+12 x^2 \log (2 x)\right)\, ,\\
\psi^{(0)}_{1, \ell=3}&=\frac{4}{15} (x-1)^2 \left(1+12 (1-5 x) x+60 x^2 (2 x-1) \log (2 x)\right)\, ,\\
\psi^{(0)}_{1, \ell=4}&=-\frac{8}{105}-\frac{32 x}{21}+\frac{1040 x^2}{49}-32 x^3+\log (2 x) \left(\frac{96 x^2}{7}-64 x^3+64 x^4\right)\, .
\end{align}
We note that in this case the integral in $r'$ in \req{psip} contains a logarithmic divergence at infinity, so that the solution in general behaves as $\psi^{(0)}_{1,\ell}\sim r^{\ell+2}\log(r/r_0)+\mathcal{O}(r^{\ell+1})$, for an unspecified scale $r_{0}$. In the solutions above we have set $r_0=M$. 

Finally, the solutions of \req{eq11} for $\ell=2,3,4$ are given by
\begin{align}\notag
\psi^{(1)}_{1,\ell=2}&=B_{2}'(x-1)^2\left(\frac{1}{3}+2 x\right)+C_{2}'\Bigg[24 (x-1)^2 x^2p(x)+\left(-2-8 x+36 x^2-24 x^3\right)\log(x)\\
&+(x-1)\left(\frac{10}{3}+\frac{56 x}{3}-24 x^2\right)\Bigg]+\frac{B_{2}}{18} (x-1)^2 (-19-18 x+6 \log (2 x) (1+6 x))\, ,\\\notag
\psi^{(1)}_{1,\ell=3}&=B_{3}'(x-1)^2\left(-\frac{1}{15}-\frac{4 x}{5}+\frac{8 x^2}{3}\right)+C_{3}'\Bigg[120 (x-1)^2 x^2 (2 x-1)p(x)\\\notag
&+\left(2+20 x-260 x^2+480 x^3-240 x^4\right)\log(x)\\\notag
&+(x-1)\left(-\frac{86}{15}-\frac{856 x}{15}+\frac{1504 x^2}{5}-240 x^3\right)\Bigg]\\
&+B_{3}(x-1)^2\left[\frac{37}{450}-\frac{76 x}{75}-\frac{8 x^2}{9}+\log (2 x) \left(-\frac{1}{15}-\frac{4 x}{5}+\frac{8 x^2}{3}\right)\right]\, ,\\\notag
\psi^{(1)}_{1,\ell=4}&=B_{4}'(x-1)^2\left(\frac{2}{105}+\frac{8 x}{21}-\frac{22 x^2}{7}+4 x^3\right)+C_{4}'\Bigg[720 (x-1)^2 x^2 \left(\frac{3}{7}+2 (x-1) x\right)p(x)\\\notag
&+\left(-\frac{12}{7}-\frac{216 x}{7}+\frac{5760 x^2}{7}-\frac{20640 x^3}{7}+3600 x^4-1440 x^5\right)\log(x)\\\notag
&+(x-1)\left(\frac{212}{35}+\frac{504 x}{5}-\frac{8328 x^2}{7}+\frac{17648 x^3}{7}-1440 x^4\right)\Bigg]\\
&+B_{4}(x-1)^2\left[-\frac{113}{22050}+\frac{358 x}{441}-\frac{115 x^2}{98}-x^3+\log (2 x) \left(\frac{2}{105}+\frac{8 x}{21}-\frac{22 x^2}{7}+4 x^3\right)\right]\, .
\end{align}
Here we have set $C_{\ell}=0$ since these solutions are only relevant in that case. We observe that the terms proportional to $B_{\ell}'$ and $C_{\ell}'$ are the same as the terms proportional to $B_{\ell}$ and $C_{\ell}$ in \req{psi12}-\req{psi14}.

	\bibliographystyle{JHEP}
	\bibliography{Gravities.bib}

\providecommand{\href}[2]{#2}\begingroup\raggedright\begin{thebibliography}{10}

\bibitem{Binnington:2009bb}
T.~Binnington and E.~Poisson, \emph{{Relativistic theory of tidal Love
  numbers}}, \href{http://dx.doi.org/10.1103/PhysRevD.80.084018}{\emph{Phys.
  Rev. D} {\bfseries 80} (2009) 084018},
  [\href{https://arxiv.org/abs/0906.1366}{{\ttfamily 0906.1366}}].

\bibitem{LIGOScientific:2016aoc}
{\scshape LIGO Scientific, Virgo} collaboration, B.~P. Abbott et~al.,
  \emph{{Observation of Gravitational Waves from a Binary Black Hole Merger}},
  \href{http://dx.doi.org/10.1103/PhysRevLett.116.061102}{\emph{Phys. Rev.
  Lett.} {\bfseries 116} (2016) 061102},
  [\href{https://arxiv.org/abs/1602.03837}{{\ttfamily 1602.03837}}].

\bibitem{Damour:2009vw}
T.~Damour and A.~Nagar, \emph{{Relativistic tidal properties of neutron
  stars}}, \href{http://dx.doi.org/10.1103/PhysRevD.80.084035}{\emph{Phys. Rev.
  D} {\bfseries 80} (2009) 084035},
  [\href{https://arxiv.org/abs/0906.0096}{{\ttfamily 0906.0096}}].

\bibitem{Vines:2011ud}
J.~Vines, E.~E. Flanagan and T.~Hinderer, \emph{{Post-1-Newtonian tidal effects
  in the gravitational waveform from binary inspirals}},
  \href{http://dx.doi.org/10.1103/PhysRevD.83.084051}{\emph{Phys. Rev. D}
  {\bfseries 83} (2011) 084051},
  [\href{https://arxiv.org/abs/1101.1673}{{\ttfamily 1101.1673}}].

\bibitem{Bini:2012gu}
D.~Bini, T.~Damour and G.~Faye, \emph{{Effective action approach to
  higher-order relativistic tidal interactions in binary systems and their
  effective one body description}},
  \href{http://dx.doi.org/10.1103/PhysRevD.85.124034}{\emph{Phys. Rev. D}
  {\bfseries 85} (2012) 124034},
  [\href{https://arxiv.org/abs/1202.3565}{{\ttfamily 1202.3565}}].

\bibitem{Cardoso:2017cfl}
V.~Cardoso, E.~Franzin, A.~Maselli, P.~Pani and G.~Raposo, \emph{{Testing
  strong-field gravity with tidal Love numbers}},
  \href{http://dx.doi.org/10.1103/PhysRevD.95.084014}{\emph{Phys. Rev. D}
  {\bfseries 95} (2017) 084014},
  [\href{https://arxiv.org/abs/1701.01116}{{\ttfamily 1701.01116}}].

\bibitem{Kol:2011vg}
B.~Kol and M.~Smolkin, \emph{{Black hole stereotyping: Induced gravito-static
  polarization}}, \href{http://dx.doi.org/10.1007/JHEP02(2012)010}{\emph{JHEP}
  {\bfseries 02} (2012) 010},
  [\href{https://arxiv.org/abs/1110.3764}{{\ttfamily 1110.3764}}].

\bibitem{Hui:2020xxx}
L.~Hui, A.~Joyce, R.~Penco, L.~Santoni and A.~R. Solomon, \emph{{Static
  response and Love numbers of Schwarzschild black holes}},
  \href{http://dx.doi.org/10.1088/1475-7516/2021/04/052}{\emph{JCAP} {\bfseries
  04} (2021) 052}, [\href{https://arxiv.org/abs/2010.00593}{{\ttfamily
  2010.00593}}].

\bibitem{LeTiec:2020bos}
A.~Le~Tiec, M.~Casals and E.~Franzin, \emph{{Tidal Love Numbers of Kerr Black
  Holes}}, \href{http://dx.doi.org/10.1103/PhysRevD.103.084021}{\emph{Phys.
  Rev. D} {\bfseries 103} (2021) 084021},
  [\href{https://arxiv.org/abs/2010.15795}{{\ttfamily 2010.15795}}].

\bibitem{Charalambous:2021mea}
P.~Charalambous, S.~Dubovsky and M.~M. Ivanov, \emph{{On the Vanishing of Love
  Numbers for Kerr Black Holes}},
  \href{http://dx.doi.org/10.1007/JHEP05(2021)038}{\emph{JHEP} {\bfseries 05}
  (2021) 038}, [\href{https://arxiv.org/abs/2102.08917}{{\ttfamily
  2102.08917}}].

\bibitem{Charalambous:2021kcz}
P.~Charalambous, S.~Dubovsky and M.~M. Ivanov, \emph{{Hidden Symmetry of
  Vanishing Love Numbers}},
  \href{http://dx.doi.org/10.1103/PhysRevLett.127.101101}{\emph{Phys. Rev.
  Lett.} {\bfseries 127} (2021) 101101},
  [\href{https://arxiv.org/abs/2103.01234}{{\ttfamily 2103.01234}}].

\bibitem{Ivanov:2022qqt}
M.~M. Ivanov and Z.~Zhou, \emph{{Vanishing of Black Hole Tidal Love Numbers
  from Scattering Amplitudes}},
  \href{http://dx.doi.org/10.1103/PhysRevLett.130.091403}{\emph{Phys. Rev.
  Lett.} {\bfseries 130} (2023) 091403},
  [\href{https://arxiv.org/abs/2209.14324}{{\ttfamily 2209.14324}}].

\bibitem{Cardoso:2018ptl}
V.~Cardoso, M.~Kimura, A.~Maselli and L.~Senatore, \emph{{Black Holes in an
  Effective Field Theory Extension of General Relativity}},
  \href{http://dx.doi.org/10.1103/PhysRevLett.121.251105}{\emph{Phys. Rev.
  Lett.} {\bfseries 121} (2018) 251105},
  [\href{https://arxiv.org/abs/1808.08962}{{\ttfamily 1808.08962}}].

\bibitem{Chakravarti:2018vlt}
K.~Chakravarti, S.~Chakraborty, S.~Bose and S.~SenGupta, \emph{{Tidal Love
  numbers of black holes and neutron stars in the presence of higher
  dimensions: Implications of GW170817}},
  \href{http://dx.doi.org/10.1103/PhysRevD.99.024036}{\emph{Phys. Rev. D}
  {\bfseries 99} (2019) 024036},
  [\href{https://arxiv.org/abs/1811.11364}{{\ttfamily 1811.11364}}].

\bibitem{Cai:2019npx}
S.~Cai and K.-D. Wang, \emph{{Non-vanishing of tidal Love numbers}},
  \href{https://arxiv.org/abs/1906.06850}{{\ttfamily 1906.06850}}.

\bibitem{DeLuca:2022tkm}
V.~De~Luca, J.~Khoury and S.~S.~C. Wong, \emph{{Implications of the weak
  gravity conjecture for tidal Love numbers of black holes}},
  \href{http://dx.doi.org/10.1103/PhysRevD.108.044066}{\emph{Phys. Rev. D}
  {\bfseries 108} (2023) 044066},
  [\href{https://arxiv.org/abs/2211.14325}{{\ttfamily 2211.14325}}].

\bibitem{Katagiri:2023umb}
T.~Katagiri, T.~Ikeda and V.~Cardoso, \emph{{Parametrized Love numbers of
  nonrotating black holes}},
  \href{http://dx.doi.org/10.1103/PhysRevD.109.044067}{\emph{Phys. Rev. D}
  {\bfseries 109} (2024) 044067},
  [\href{https://arxiv.org/abs/2310.19705}{{\ttfamily 2310.19705}}].

\bibitem{Katagiri:2024fpn}
T.~Katagiri, V.~Cardoso, T.~Ikeda and K.~Yagi, \emph{{Tidal response beyond
  vacuum General Relativity with a canonical definition}},
  \href{https://arxiv.org/abs/2410.02531}{{\ttfamily 2410.02531}}.

\bibitem{Chakraborty:2024gcr}
S.~Chakraborty, G.~Comp\`ere and L.~Machet, \emph{{Tidal Love numbers and
  quasi-normal modes of the Schwarzschild-Hernquist black hole}},
  \href{https://arxiv.org/abs/2412.14831}{{\ttfamily 2412.14831}}.

\bibitem{Barbosa:2025uau}
S.~Barbosa, P.~Brax, S.~Fichet and L.~de~Souza, \emph{{Running Love Numbers and
  the Effective Field Theory of Gravity}},
  \href{https://arxiv.org/abs/2501.18684}{{\ttfamily 2501.18684}}.

\bibitem{Yazadjiev:2018xxk}
S.~S. Yazadjiev, D.~D. Doneva and K.~D. Kokkotas, \emph{{Tidal Love numbers of
  neutron stars in $f(R)$ gravity}},
  \href{http://dx.doi.org/10.1140/epjc/s10052-018-6285-z}{\emph{Eur. Phys. J.
  C} {\bfseries 78} (2018) 818},
  [\href{https://arxiv.org/abs/1803.09534}{{\ttfamily 1803.09534}}].

\bibitem{Diedrichs:2025vhv}
R.~F. Diedrichs, S.~Tsujikawa and K.~Yagi, \emph{{Tidal Love Numbers of Neutron
  Stars in Horndeski Theories}},
  \href{https://arxiv.org/abs/2501.07998}{{\ttfamily 2501.07998}}.

\bibitem{regge1957stability}
T.~Regge and J.~A. Wheeler, \emph{Stability of a schwarzschild singularity},
  {\emph{Physical Review} {\bfseries 108} (1957) 1063}.

\bibitem{zerilli1970effective}
F.~J. Zerilli, \emph{Effective potential for even-parity regge-wheeler
  gravitational perturbation equations}, {\emph{Physical Review Letters}
  {\bfseries 24} (1970) 737}.

\bibitem{LeTiec:2020spy}
A.~Le~Tiec and M.~Casals, \emph{{Spinning Black Holes Fall in Love}},
  \href{http://dx.doi.org/10.1103/PhysRevLett.126.131102}{\emph{Phys. Rev.
  Lett.} {\bfseries 126} (2021) 131102},
  [\href{https://arxiv.org/abs/2007.00214}{{\ttfamily 2007.00214}}].

\bibitem{Chia:2020yla}
H.~S. Chia, \emph{{Tidal deformation and dissipation of rotating black holes}},
  \href{http://dx.doi.org/10.1103/PhysRevD.104.024013}{\emph{Phys. Rev. D}
  {\bfseries 104} (2021) 024013},
  [\href{https://arxiv.org/abs/2010.07300}{{\ttfamily 2010.07300}}].

\bibitem{Rodriguez:2023xjd}
M.~J. Rodriguez, L.~Santoni, A.~R. Solomon and L.~F. Temoche, \emph{{Love
  numbers for rotating black holes in higher dimensions}},
  \href{http://dx.doi.org/10.1103/PhysRevD.108.084011}{\emph{Phys. Rev. D}
  {\bfseries 108} (2023) 084011},
  [\href{https://arxiv.org/abs/2304.03743}{{\ttfamily 2304.03743}}].

\bibitem{Perry:2023wmm}
M.~Perry and M.~J. Rodriguez, \emph{{Dynamical Love Numbers for Kerr Black
  Holes}},  \href{https://arxiv.org/abs/2310.03660}{{\ttfamily 2310.03660}}.

\bibitem{Perry:2024vwz}
M.~Perry and M.~J. Rodriguez, \emph{{Love Numbers for Extremal Kerr Black
  Hole}},  \href{https://arxiv.org/abs/2412.19699}{{\ttfamily 2412.19699}}.

\bibitem{Bhatt:2024yyz}
R.~P. Bhatt, S.~Chakraborty and S.~Bose, \emph{{Rotating black holes experience
  dynamical tides}},
  \href{http://dx.doi.org/10.1103/PhysRevD.111.L041504}{\emph{Phys. Rev. D}
  {\bfseries 111} (2025) L041504},
  [\href{https://arxiv.org/abs/2406.09543}{{\ttfamily 2406.09543}}].

\bibitem{Teukolsky:1972my}
S.~A. Teukolsky, \emph{{Rotating black holes - separable wave equations for
  gravitational and electromagnetic perturbations}},
  \href{http://dx.doi.org/10.1103/PhysRevLett.29.1114}{\emph{Phys. Rev. Lett.}
  {\bfseries 29} (1972) 1114--1118}.

\bibitem{Teukolsky:1973ha}
S.~A. Teukolsky, \emph{{Perturbations of a rotating black hole. 1. Fundamental
  equations for gravitational electromagnetic and neutrino field
  perturbations}}, \href{http://dx.doi.org/10.1086/152444}{\emph{Astrophys. J.}
  {\bfseries 185} (1973) 635--647}.

\bibitem{Li:2022pcy}
D.~Li, P.~Wagle, Y.~Chen and N.~Yunes, \emph{{Perturbations of Spinning Black
  Holes beyond General Relativity: Modified Teukolsky Equation}},
  \href{http://dx.doi.org/10.1103/PhysRevX.13.021029}{\emph{Phys. Rev. X}
  {\bfseries 13} (2023) 021029},
  [\href{https://arxiv.org/abs/2206.10652}{{\ttfamily 2206.10652}}].

\bibitem{Hussain:2022ins}
A.~Hussain and A.~Zimmerman, \emph{{Approach to computing spectral shifts for
  black holes beyond Kerr}},
  \href{http://dx.doi.org/10.1103/PhysRevD.106.104018}{\emph{Phys. Rev. D}
  {\bfseries 106} (2022) 104018},
  [\href{https://arxiv.org/abs/2206.10653}{{\ttfamily 2206.10653}}].

\bibitem{Cano:2023tmv}
P.~A. Cano, K.~Fransen, T.~Hertog and S.~Maenaut, \emph{{Universal Teukolsky
  equations and black hole perturbations in higher-derivative gravity}},
  \href{http://dx.doi.org/10.1103/PhysRevD.108.024040}{\emph{Phys. Rev. D}
  {\bfseries 108} (2023) 024040},
  [\href{https://arxiv.org/abs/2304.02663}{{\ttfamily 2304.02663}}].

\bibitem{Cano:2023jbk}
P.~A. Cano, K.~Fransen, T.~Hertog and S.~Maenaut, \emph{{Quasinormal modes of
  rotating black holes in higher-derivative gravity}},
  \href{http://dx.doi.org/10.1103/PhysRevD.108.124032}{\emph{Phys. Rev. D}
  {\bfseries 108} (2023) 124032},
  [\href{https://arxiv.org/abs/2307.07431}{{\ttfamily 2307.07431}}].

\bibitem{Wagle:2023fwl}
P.~Wagle, D.~Li, Y.~Chen and N.~Yunes, \emph{{Perturbations of spinning black
  holes in dynamical Chern-Simons gravity: Slow rotation equations}},
  \href{http://dx.doi.org/10.1103/PhysRevD.109.104029}{\emph{Phys. Rev. D}
  {\bfseries 109} (2024) 104029},
  [\href{https://arxiv.org/abs/2311.07706}{{\ttfamily 2311.07706}}].

\bibitem{Cano:2024ezp}
P.~A. Cano, L.~Capuano, N.~Franchini, S.~Maenaut and S.~H. V\"olkel,
  \emph{{Higher-derivative corrections to the Kerr quasinormal mode spectrum}},
  \href{http://dx.doi.org/10.1103/PhysRevD.110.124057}{\emph{Phys. Rev. D}
  {\bfseries 110} (2024) 124057},
  [\href{https://arxiv.org/abs/2409.04517}{{\ttfamily 2409.04517}}].

\bibitem{Guo:2024bqe}
R.-Z. Guo, H.~Tan and Q.-G. Huang, \emph{{Generic Modified Teukolsky Formalism
  beyond General Relativity for Spherically Symmetric Cases}},
  \href{https://arxiv.org/abs/2409.14437}{{\ttfamily 2409.14437}}.

\bibitem{Gralla:2017djj}
S.~E. Gralla, \emph{{On the Ambiguity in Relativistic Tidal Deformability}},
  \href{http://dx.doi.org/10.1088/1361-6382/aab186}{\emph{Class. Quant. Grav.}
  {\bfseries 35} (2018) 085002},
  [\href{https://arxiv.org/abs/1710.11096}{{\ttfamily 1710.11096}}].

\bibitem{Ivanov:2022hlo}
M.~M. Ivanov and Z.~Zhou, \emph{{Revisiting the matching of black hole tidal
  responses: A systematic study of relativistic and logarithmic corrections}},
  \href{http://dx.doi.org/10.1103/PhysRevD.107.084030}{\emph{Phys. Rev. D}
  {\bfseries 107} (2023) 084030},
  [\href{https://arxiv.org/abs/2208.08459}{{\ttfamily 2208.08459}}].

\bibitem{Bhatt:2023zsy}
R.~P. Bhatt, S.~Chakraborty and S.~Bose, \emph{{Addressing issues in defining
  the Love numbers for black holes}},
  \href{http://dx.doi.org/10.1103/PhysRevD.108.084013}{\emph{Phys. Rev. D}
  {\bfseries 108} (2023) 084013},
  [\href{https://arxiv.org/abs/2306.13627}{{\ttfamily 2306.13627}}].

\bibitem{Goldberger:2004jt}
W.~D. Goldberger and I.~Z. Rothstein, \emph{{An Effective field theory of
  gravity for extended objects}},
  \href{http://dx.doi.org/10.1103/PhysRevD.73.104029}{\emph{Phys. Rev. D}
  {\bfseries 73} (2006) 104029},
  [\href{https://arxiv.org/abs/hep-th/0409156}{{\ttfamily hep-th/0409156}}].

\bibitem{Porto:2005ac}
R.~A. Porto, \emph{{Post-Newtonian corrections to the motion of spinning bodies
  in NRGR}}, \href{http://dx.doi.org/10.1103/PhysRevD.73.104031}{\emph{Phys.
  Rev. D} {\bfseries 73} (2006) 104031},
  [\href{https://arxiv.org/abs/gr-qc/0511061}{{\ttfamily gr-qc/0511061}}].

\bibitem{Levi:2018nxp}
M.~Levi, \emph{{Effective Field Theories of Post-Newtonian Gravity: A
  comprehensive review}},
  \href{http://dx.doi.org/10.1088/1361-6633/ab12bc}{\emph{Rept. Prog. Phys.}
  {\bfseries 83} (2020) 075901},
  [\href{https://arxiv.org/abs/1807.01699}{{\ttfamily 1807.01699}}].

\bibitem{Saketh:2023bul}
M.~V.~S. Saketh, Z.~Zhou and M.~M. Ivanov, \emph{{Dynamical tidal response of
  Kerr black holes from scattering amplitudes}},
  \href{http://dx.doi.org/10.1103/PhysRevD.109.064058}{\emph{Phys. Rev. D}
  {\bfseries 109} (2024) 064058},
  [\href{https://arxiv.org/abs/2307.10391}{{\ttfamily 2307.10391}}].

\bibitem{Endlich:2017tqa}
S.~Endlich, V.~Gorbenko, J.~Huang and L.~Senatore, \emph{{An effective
  formalism for testing extensions to General Relativity with gravitational
  waves}}, \href{http://dx.doi.org/10.1007/JHEP09(2017)122}{\emph{JHEP}
  {\bfseries 09} (2017) 122},
  [\href{https://arxiv.org/abs/1704.01590}{{\ttfamily 1704.01590}}].

\bibitem{Cano:2019ore}
P.~A. Cano and A.~Ruip\'erez, \emph{{Leading higher-derivative corrections to
  Kerr geometry}}, \href{http://dx.doi.org/10.1007/JHEP05(2019)189}{\emph{JHEP}
  {\bfseries 05} (2019) 189},
  [\href{https://arxiv.org/abs/1901.01315}{{\ttfamily 1901.01315}}].

\bibitem{Cano:2020cao}
P.~A. Cano, K.~Fransen and T.~Hertog, \emph{{Ringing of rotating black holes in
  higher-derivative gravity}},
  \href{http://dx.doi.org/10.1103/PhysRevD.102.044047}{\emph{Phys. Rev. D}
  {\bfseries 102} (2020) 044047},
  [\href{https://arxiv.org/abs/2005.03671}{{\ttfamily 2005.03671}}].

\bibitem{deRham:2020ejn}
C.~de~Rham, J.~Francfort and J.~Zhang, \emph{{Black Hole Gravitational Waves in
  the Effective Field Theory of Gravity}},
  \href{http://dx.doi.org/10.1103/PhysRevD.102.024079}{\emph{Phys. Rev. D}
  {\bfseries 102} (2020) 024079},
  [\href{https://arxiv.org/abs/2005.13923}{{\ttfamily 2005.13923}}].

\bibitem{Cano:2021myl}
P.~A. Cano, K.~Fransen, T.~Hertog and S.~Maenaut, \emph{{Gravitational ringing
  of rotating black holes in higher-derivative gravity}},
  \href{http://dx.doi.org/10.1103/PhysRevD.105.024064}{\emph{Phys. Rev. D}
  {\bfseries 105} (2022) 024064},
  [\href{https://arxiv.org/abs/2110.11378}{{\ttfamily 2110.11378}}].

\bibitem{gitbeyondkerr}
P.~A. Cano, \emph{{https://github.com/pacmn91/BeyondKerrQNM}},  2025.

\bibitem{Goldberger:2020fot}
W.~D. Goldberger, J.~Li and I.~Z. Rothstein, \emph{{Non-conservative effects on
  spinning black holes from world-line effective field theory}},
  \href{http://dx.doi.org/10.1007/JHEP06(2021)053}{\emph{JHEP} {\bfseries 06}
  (2021) 053}, [\href{https://arxiv.org/abs/2012.14869}{{\ttfamily
  2012.14869}}].

\end{thebibliography}\endgroup
	
\end{document}